\documentclass[structabstract]{aa}  
\usepackage{graphicx}
\usepackage{natbib}
\usepackage{color}
\usepackage{multirow}
\usepackage{amssymb,amsmath}
\bibpunct{(}{)}{;}{a}{}{,} 

\def\vslip{$v_{\mathrm{slip}}$}
\def\vpeak{$v_{\mathrm{peak}}$}
\def\vQSL{$v_{\mathrm{QSL}}$}
\def\cA{$\overline{c_{A}}$}
\def\Npeak{$N_{\mathrm{peak}}$}
\def\Bratio{$B_z^{\mathrm{ratio}}$}
\def\Qroot{$Q^{1/2}$}
\def\ie{\textit{i.e.}}
\defcitealias{Aulanier12}{Paper~I}

\begin{document}
   \title{The standard flare model in three dimensions}

   \subtitle{III. Slip-running reconnection properties}

   \author{M. Janvier 
          \and
          G. Aulanier   
          \and
          E. Pariat 
          \and
          P. D\'emoulin 
          }

   \institute{LESIA, Observatoire de Paris, CNRS, UPMC, Univ. Paris Diderot, 5 place Jules Janssen, 92190 Meudon\\
              \email{miho.janvier@obspm.fr}
             }

   \date{Received ... ; accepted ...}


  \abstract
    {A standard model for eruptive flares aims at describing observational 3D features of the reconnecting coronal magnetic field. Extensions to the 2D model require the physical understanding of 3D reconnection processes at the origin of the magnetic configuration evolution. However, the properties of 3D reconnection without null point and separatrices still need to be analyzed.}
   {We focus on magnetic reconnection associated with the growth and evolution of a flux rope and associated flare loops during an eruptive flare. We aim at understanding the intrinsic characteristics of 3D reconnection in the presence of quasi-separatrix layers (QSLs), how QSL properties are related to the slip-running reconnection mode in general, and how this applies to eruptive flares in particular.}
   {We studied the slip-running reconnection of field lines in a magnetohydrodynamic simulation of an eruptive flare associated with a torus-unstable flux rope. The squashing degree and the mapping norm are two parameters related to the QSLs. We computed them to investigate their relation with the slip-running reconnection speed of selected field lines.} 
   {Field lines associated with the flux rope and the flare loops undergo a continuous series of magnetic reconnection, which results in their super-Alfv\'enic slipping motion. The time profile of their slippage speed and the space distribution of the mapping norm are shown to be strongly correlated. We find that the motion speed is proportional to the mapping norm. Moreover, this slip-running motion becomes faster as the flux rope expands, since the 3D current layer evolves toward a current sheet, and QSLs to separatrices.
   } 
   {The present analysis extends our understanding of the 3D slip-running reconnection regime. We identified a controlling parameter of the apparent velocity of field lines while they slip-reconnect, enabling the interpretation of the evolution of post flare loops. This work completes the standard model for flares and 
eruptions by giving its 3D properties.
}

   \maketitle


\section{Introduction}
\label{secintro}

Eruptive flares are major energetic events taking place in the Sun's atmosphere that extend to the interplanetary medium, and especially affect the near-Earth environment \citep[\textit{e.g.}][]{Gosling91}. {Their evolution is associated} with the formation of the coronal mass ejection (CME) and flare loops (also defined as post-flare loops), {and their mechanism has long been explained} with a simplified model depicting reconnecting coronal loops. This model, also referred to as the CSHKP model \citep{Carmichael64,Sturrock66,Hirayama74,Kopp76}, captures the main features of observed signatures of eruptive flares but also of confined flares (\ie, not associated with a CME). Among these signatures are the formation of the flux rope \citep[\textit{e.g.}][]{Sakurai76,Dere99,Cheng11,Cheng13} as well as the flare loops and the flare ribbons \citep[\textit{e.g.}][]{Schmieder96}, correlated with the impact of high-energy particles with the chromosphere \citep[\textit{e.g.}][]{Reid12}.

Magnetic reconnection is believed to be the core process at work in these events, with numerous observational evidence \citep[\textit{e.g.}][]{Tsuneta92,Schmieder97,Yokoyama01,McKenzie11}. This process releases the energy stored in the coronal magnetic field by converting it into heating and kinetic energy, and by injecting energetic particles into reconnected loops \citep[\textit{e.g.}][]{Aschwanden96} and the CME \citep[\textit{e.g.}][]{Masson09,Masson12}.

Over the years, extended observations of eruptive flares during their evolution have revealed typical 3D magnetic structures. Among them, twisted flux ropes often appear in pre-flare/flaring regions \citep[\textit{e.g.}][]{Canou10,Guo12}. $J-$ and $S-$ shaped sigmoids \citep{Gibson02,Savcheva09, Green11, Savcheva12} as well as sheared flare loops \citep{Asai03, Warren11, Aulanier12} are commonly observed.
Since the CSHKP model is restricted in terms of 3D features intrinsic to flares, 3D extensions have been proposed to explain the expansion of the flux rope associated with the CME.
Models inspired by observations and developed numerically include reconnection of sheared arcades, also referred to as tether-cutting \citep[][]{Moore97,Fan12}, formation and evolution of twisted flux ropes \citep{Shibata95,Shiota05,Priest02,Amari03a,Amari03b}, and unstable flux ropes \citep{Torok04,Aulanier10}.

The 3D features of an eruptive flare were studied in detail in the first paper of the present series \citep{Aulanier12}, hereafter referred to as \citetalias{Aulanier12}. There, the authors investigated the changes in the shear during the formation and evolution of flare loops. A 3D flux rope simulation was used to investigate the May 9, 2011 event (see also \citealp{Warren11}). {It was shown} that the shear found in flare loops is transferred from the pre-eruptive expanding magnetic field via {the gradual formation of the loops by reconnection}.
In the present paper, which is the third of the series, 3D reconnection, {\ie, the mechanism that forms} both flux rope and flare loops, is studied in detail. 

Reconnection corresponds to the change of magnetic {connectivity}, \ie, when field lines ``break'' and reconnect with each other. This process exists only in regions where the ideal MHD breaks down, \ie, when the plasma frozen-in condition ceases to be valid. A steady-state mechanism explaining the release of magnetic energy was first given in \citealt{Sweet58,Parker57}, with extensions more recently in 3D \citep[\textit{e.g.}][]{Baty00}. In this model, a small diffusion region exists where the plasma is no longer ``attached'' to the magnetic field, allowing new connections of magnetic field lines and defining four domains of magnetic connectivity. The frontiers of different {domains} are introduced as separatrices, while their intersection is referred to as a null point, which can exist both in 2D an 3D \citep{Lau90}. 
However, although they necessarily exist in 2D reconnection, 3D reconnecting configurations do not always require separatrices and null points \citep[see][and references therein]{Longcope05}.

\citet{Demoulin96a, Demoulin97}, for example, showed from linear force-free magnetic field extrapolations that the photospheric footpoint mapping can be continuous (\ie, without separatrices) in regions associated with solar flare events. In such a configuration, domains corresponding to drastic changes in the magnetic field connectivity gradient are identified as quasi-separatrix layers (QSLs). {The central part of these volumes corresponds to the strongest field line distortion and is defined as a hyperbolic flux tube (HFT, \citealp{Titov02}).}
QSLs and especially HFTs have been shown to be associated with locations of high electric current density regions (\textit{e.g.} \citealp{Aulanier05b,Masson09,Wilmot09}). 
The latter behave similarly to current sheets associated with separatrices, because ideal MHD can break down and magnetic reconnection can take place \citep{Aulanier06}. 
In such cases, magnetic field lines ``slip'' or ``flip'' inside the plasma as suggested by \citet{Priest95} and subsequently investigated in \citet{Priest03}. This 3D reconnecting mode explains observations of coronal loop motion with the Hinode spacecraft \citep{Aulanier07}. {Finally, \citet{Masson12} showed that a model involving slipping reconnection can explain} interchange reconnection between close and open field lines in the corona, allowing the injection of energetic particles into the interplanetary space.

Depending on the speed of the flipping motion, two definitions were given in \citet{Aulanier06}: if the speed is sub-Alfv\'enic (\vslip$< c_A$), the field lines have a \textit{slipping} motion, while if the speed is super-alfv\'enic, the apparent motion of the reconnecting field lines is said to be due to \textit{slip-running} reconnection.
{The authors also investigated the properties of the slipping motion and the possible relation with that of the QSLs. Although it was discussed that the slipping motion should have a velocity proportional to \Qroot, where $Q$ stands for the squashing degree defining QSLs, a strict correlation study was not explicitly achieved.} 

In the present paper, we investigate in detail the features of the reconnection process that leads to the formation of flare loops during the flux rope ejection. 
Especially, we extensively use parameters that define QSLs, such as the norm of the connectivity mapping $N$ \citep{Priest95} and the squashing degree $Q$ \citep{Titov02}, to find possible relations with the speed of the slipping motion of field lines.
Thus, understanding how magnetic reconnection works in the presence of QSLs provides a generalization of the mechanisms in the 3D framework. This knowledge is important for understanding observational features of solar reconnection events and for extending the standard flare model in 3D. In particular, we explain the formation of the flux rope envelope and the flare loops, as well as the flare ribbons and chromospheric footpoint motions with the QSL reconnection model.

The structure of the paper is as follows: Section \ref{sec2} presents details of the MHD simulation of the flux rope ejection and the general evolution of the magnetic structure, its topology, and the associated electric currents. Section \ref{sec3} focuses on the formation of flare loops and the subsequent build-up of the flux rope via slip-running reconnection, and their relation with the spatial distribution of QSLs. Section \ref{sec4} correlates the slip-running reconnection speed with QSL parameters, and characterizes the evolution of QSLs throughout the simulation. Finally, the results are summarized in Sect. \ref{secsum} and are discussed along with concluding remarks in Sect.  \ref{secccl}.

	\begin{figure*}
     	\centering
	\sidecaption
    	\includegraphics[bb=10 128 571 700,width=12cm,clip]{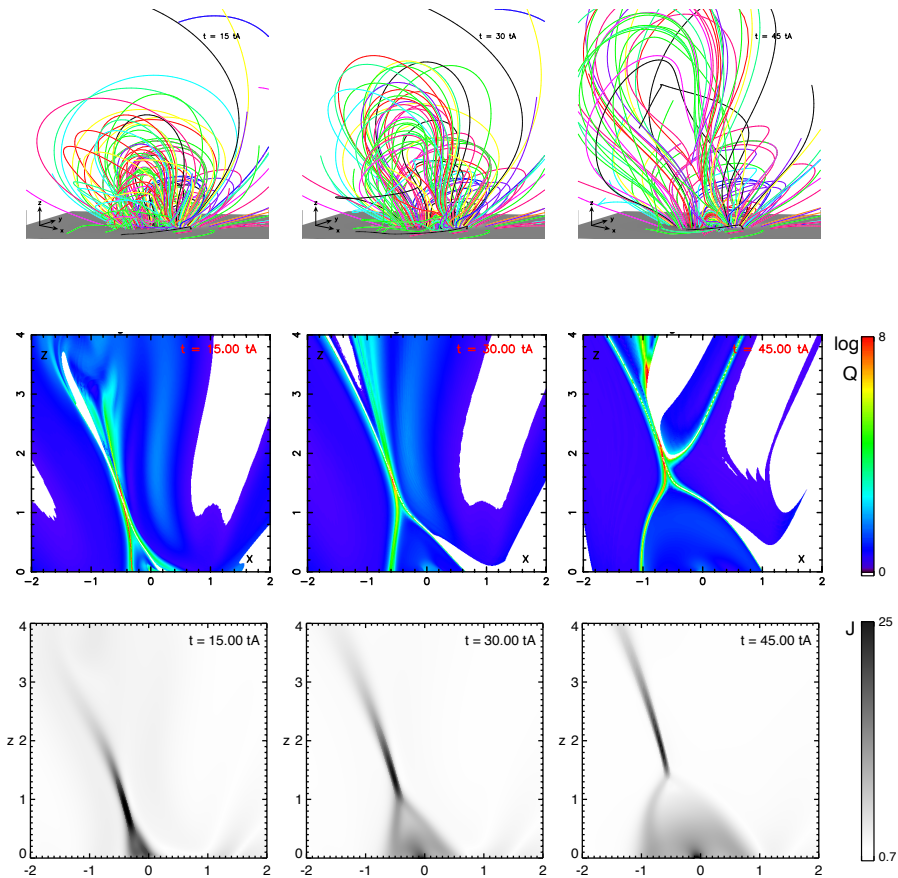}
	   	\caption{Global evolution of the magnetic field lines, the QSL, and the electric current during the flux rope ejection. \textbf{Top row}: sets of random field lines at $t=15, 30,$ and $45\ t_A$. Their colors are chosen randomly. \textbf{Middle row}: 2D vertical cuts for the decimal logarithm of the squashing degree $Q$ for $y=-0.3$. High values of $Q$ are red-colored. White corresponds to magnetic field lines reaching the {sides of} the numerical box.
		\textbf{Bottom row}: Corresponding vertical cuts of the current density $J(x,-0.3,z)$. 
		     	}
     	\label{fig1}
     	\end{figure*}


\section{MHD simulation of an eruptive flare}
\label{sec2}

\subsection{Initial conditions, equations, and numerical domain}
\label{subsec21}

We used a 3D MHD simulation {that reproduces} the ejection of a flux rope during an eruptive flare (identical to \citetalias{Aulanier12}). A twisted flux rope is first constructed from shearing motions and diffusion at the photosphere of a bipolar and asymmetric potential magnetic field. As shown in \citet{Aulanier10}, a threshold is reached at a critical step of the evolution, compatible with the onset of the torus instability. In the present simulation, the {height} of the magnetic flux rope was slightly above this threshold and is provided as an initial condition. No external forcing was applied, so that the torus instability leads to the subsequent expansion and ejection of the flux rope accompanied by reconnecting loops in the wake of the simulated CME.

The 3D numerical simulation was performed with the \textit{observationally-driven high-order scheme magneto-hydrodynamic} code (OHM, \citealp{Aulanier05a}). This solves the full MHD equations for the mass density $\rho$, the fluid velocity $\vec{u}$, and the magnetic field $\vec{B}$ under the plasma $\beta$=0 assumption. All parameters are expressed in non-dimensionalized units, with the averaged Alfv\'en speed \cA=1 used to normalize the velocities, and $t_A=1$ the travel time for a distance $d=1$ at \cA for time normalization. Typically, the whole size of the photospheric bipole is $L^{\mathrm{dipole}}=5$ \citep[see][]{Aulanier13}.

The time-dependent parameters are advected in Cartesian coordinates in the MHD equations:
\\
\\
\begin{eqnarray}
\label{eqcont}
 \frac{\partial \rho}{\partial t} &=& - {\vec \nabla} \cdot (\rho {\vec u}) 
 + \zeta\, \Delta (\rho-\rho_\circ)\\
\label{eqmom}
 \frac{\partial {\vec u}}{\partial t} &=& - ({\vec u} \cdot
          {\vec \nabla}) {\vec u} + ({\vec \nabla} \times {\vec B}) \times {\vec B}/(\mu\rho)
                    + \tilde{\nu}\, \tilde{\Delta} {\vec u} \\
\label{eqinduc}
 \frac{\partial {\vec B}}{\partial t} &=& {\vec \nabla} \times
         ({\vec u} \times {\vec B}) + \eta\, \Delta {\vec B} \, ,
\end{eqnarray}
where $\eta$, $\xi$ and $\tilde{\nu}$ are diffusive coefficients. In particular, $\eta$ stands for the  magnetic diffusivity responsible for the reconnection process and is set to be constant throughout the whole domain except at the photospheric boundary at $z=0$, where $\eta=0$. {Line-tied conditions are set for $z=0$, while} the other five sides of the simulation box have open boundaries. The simulation domain extends to $[-10,10]$ in the $(x;y)$ plane and to $[0,30]$ in the $z$-direction, with a non-uniform mesh. {The velocity is initially set to 0 in the whole volume, but because the system is in the torus-unstable domain, }the evolution of the flux rope is driven by the magnetic forces within the volume. More detailed descriptions of the numerical settings are given in section 3.3 and Table 1 of \citetalias{Aulanier12}.

The evolution of a set of random field lines representing the coronal magnetic field is given in the top row of  Fig.~\ref{fig1}. These field lines were drawn with the TOPOTR package \citep{Demoulin96a} and their colors chosen randomly to distinguish line bundles.
The selected times show the outward expansion of $\Omega$-shaped field lines. They reconnect at low height, additionally building up the flux rope, which was already partly formed in the pre-eruptive phase \citep[see][]{Aulanier10}. Below, arch-shaped flare loops form in the center of the simulation box in the wake of the CME.
The narrow field structure just above the flare loops indicates the location of the thin current layer where reconnection takes place.

\subsection{Coronal magnetic topology and electric currents}
\label{subsec22}

In the absence of separatrices and null points, magnetic field reconnection in the solar corona is believed to take place in QSLs \citep[][]{Demoulin97}. These are regions of high distortion of the mapping of magnetic field lines anchored in the photosphere \citep{Priest95,Demoulin96b}. This distortion is described by derivatives of the field line mapping functions expressed via Jacobian matrices, defined in the positive and the negative photospheric regions. The first definition of a QSL was given by means of the norm $N$ of the Jacobi matrix (Eq.[1] in \citealp{Demoulin96a}), also referred to as the mapping norm. Another definition, the degree of the squashing of the mapping $Q$, has been given in \citet{Titov02}. Naming the ratio of the vertical components of the magnetic field $B_z^{\mathrm{ratio}}$ at each footpoint, the squashing degree $Q$ is related to $N$ as $Q = N^2/B_z^{\mathrm{ratio}}$.

	\begin{figure}
    	\centering
         \includegraphics[bb=25 385 605 670,width=0.450\textwidth,clip]{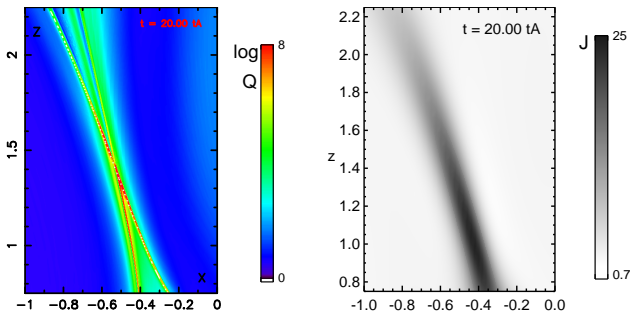}
   	\caption{Zoom of the 2D vertical cut at $t=20\ t_A$ and $y=-0.3$ comparing (\textbf{left}) the logarithm of the squashing degree $Q$ and (\textbf{right}) the current density $J$. The gray/color scales are the same as in Fig. \ref{fig1}. 
		}
	\label{fig2}
     	\end{figure}

 	\begin{figure*}
    	\centering
         \includegraphics[bb=30 550 620 750,width=1\textwidth,clip]{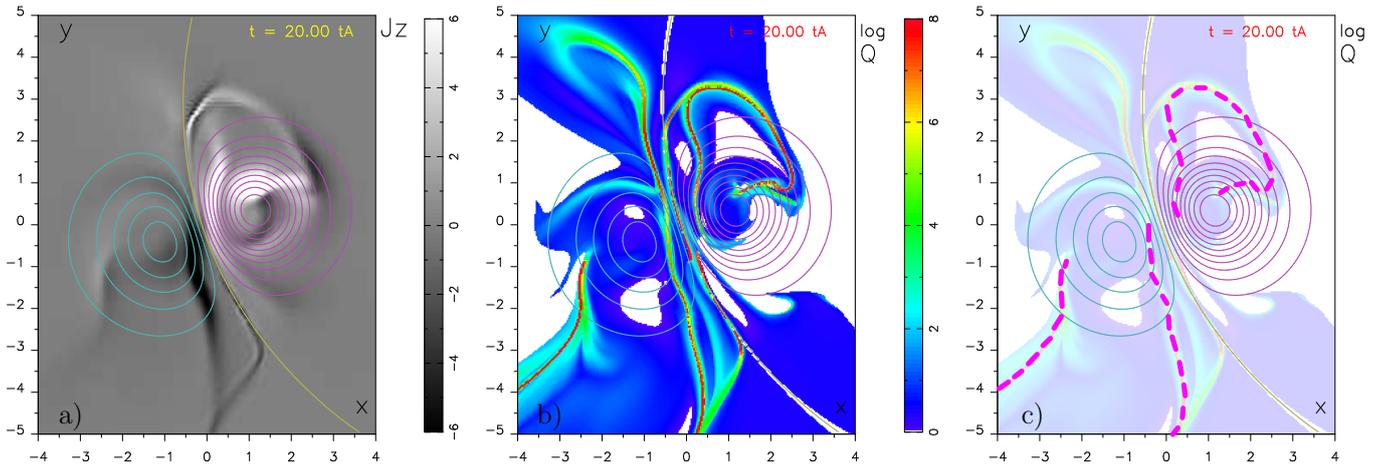}
   	\caption{Comparison of electric currents, magnetic field, and Q distribution at $z=0$. \textbf{a)} Plots of the photospheric current $J_z$ (gray-scale image) and magnetic field $B_z(z=0)$ (cyan/pink overplotted contours for the negative/positive values, respectively). Both direct and return currents exist in the two magnetic polarities (separated by the photospheric inversion line shown in yellow). \textbf{b)} Logarithm of the squashing degree $Q(z=0)$ at $t = 20\ t_A$ showing a similar double-J structure. The color-coding for $\log Q$ is the same as in Fig.  \ref{fig1}. \textbf{c)} The footprints of the main QSLs, related with the formation of the flux rope and the flare loops, are highlighted with thick dashed pink lines {on top of the shaded $Q$-map of panel b)}.
   	}
	\label{fig3}
     	\end{figure*}

Using TOPOTR, we calculated the squashing degree $Q$ and plot the related QSLs (details of these calculations can be found in \citealp{Pariat12}). Vertical 2D cuts of the QSLs showing $\log Q(x,z)$ at $y=-0.3$  are presented in the middle row of Fig. \ref{fig1}. The $y$-position has been chosen so that it cuts the flux rope in its center.
Note that these vertical cuts are only shown in $x \in [-2;2], z \in [0;4]$. This is the region where flare loops are formed, as discussed in the description of the top row panels.

These cuts reveal the time-evolution of the region surrounding the HFT, the thin volume where $Q$ reaches its highest values \citep{Titov02} and where reconnection can take place. {This coronal portion of the HFT shows a typical X-shape corresponding to a vertical cross-section of the QSL volume (Fig. \ref{fig1}).}
At $t=15\ t_A$, {the 2D~cut reveals a tear-drop shape in the top-left corner, identifying the envelope of the flux rope \citep[similarly to Fig.~8 in][]{Savcheva12}}. The upward motion {associated with its ejection} can be seen from the changes at $t=30$ and $45 \ t_A$. 
{A cusp also forms underneath the flux rope envelope and the legs of the cusp are located around the arch-shaped flare loops found in the center of the simulation box (see top row of Fig. \ref{fig1}}).
Note that the white areas in these cuts correspond to regions where the squashing degree $Q$ could not be calculated. Indeed, some magnetic field lines go beyond the simulation box, so that the integration along those lines down to the lower boundary and therefore the calculation of $Q$ is irrelevant.

Co-temporal and co-spatial 2D vertical cuts of the electric current $J(x,z)$ at $y=-0.3$ are shown in the bottom row of Fig.  \ref{fig1}. From $t=15\ t_A$ to $t=45\ t_A$, {these panels} show the upward motion and the thinning of the vertical current layer where reconnection of field lines takes place. The electric current presents an inverse-Y configuration and the cusp has the same location as that of the HFT.
This cusp also surrounds the arch-shaped flare loops and corresponds to the boundary between pre- and post-reconnected field lines.

Figure \ref{fig2} shows a zoom of the 2D vertical cut for both the HFT and the current layer from Fig. \ref{fig1} at $t=20\ t_A$. The X-shape typical of the QSL cross-section in the corona is seen (about mid-way from the photospheric footpoints, see also \citealp{Titov02} and figure 6 in \citealp{Aulanier05b}). Thin branches can also be seen in yellow-green colors at the side of the central cross shape. They demonstrate a complex QSL/HFT structure associated with the present magnetic configuration, which has more structures than a ``mere'' X-shaped crossing.
Since the HFT corresponds to the high-$Q$ region, and since intense currents build up where field lines anchored very close to each other have opposite feet widely separated \citep{Demoulin96a}, one can expect the formation of a high-current region around the HFT, such as found in \citet{Aulanier05b} and \citet{Effenberger11} in non-eruptive models. Similar results are obtained here, as shown with the zoom of the current layer on the right panel of Fig. \ref{fig2}, as the intense current layer exists where the HFT is localized. Note that in the 2D cuts of QSLs at $t=45\ t_A$, a QSL branch not associated to any current density structure appears. This QSL does not have a high intensity and is related to a region where magnetic field lines are branched out of the simulation box ($Q$ therefore corresponds to a different reference boundary and its calculation is therefore irrelevant). This shows that QSLs can exist without being necessarily related with current densities. 

\subsection{Photospheric QSL traces and electric currents}
\label{subsec23}

Figure \ref{fig3} shows top views of the photospheric current $J_z$ and the logarithm of the squashing degree $\log Q$ on the photosphere at $z=0$, $t=20\ t_A$.

In Figure \ref{fig3}a, the two magnetic polarities show several electric current patterns. Two patches of strong direct currents ($J_z/B_z>0$) can be found in the center of the two magnetic polarities, as well as small patches of weak return currents ($J_z/B_z<0$). Note that in the present paper, the term ``direct currents'' should not be confused with DC currents used to describe currents with transit times longer than the Alfv\'en transit time in the corona, which appear in coronal heating studies. In the positive polarity (pink isocontours), one of the direct current patches ($J_z>0$) starts in the center of the polarity and extends following a hook shape in the higher $y$-value region. The second direct current patch is located very near the photospheric inversion line (PIL).
Both direct currents have a counterpart in the negative polarity (where direct currents stand for $J_z<0$): elongated patches of strong direct current extend toward the lower $y$-value region along the PIL and there is a patch of weak return current near the center of the negative polarity, although not as strong as the patch in the positive polarity. These differences are due to the choice of an asymmetric initial magnetic configuration, as discussed in \citet{Aulanier13}.

The long and thin patches of weak direct current along the PIL correspond to a bald patch region, but this is not involved in the flare. Next, the narrow current layers of strong direct currents near the PIL are parallel to each other and form two current ribbons that are strongly similar to flare ribbons as observed, for example, in \citet{Fletcher01}. These current ribbons are associated with flare loops that form during the ejection of the flux rope.
 
The QSL mapping onto the photospheric plane (see Fig.~\ref{fig3}b) reveals similar structures as the electric current. The intense $Q$-layers (in red) present a hook shape similar to that of the current density. We find in this map that some structures seen in the QSL at the photosphere do not have an equivalent in the current-density map of Fig.~\ref{fig3}a. For example, the QSL extending in the negative-$x$/positive-$y$ region does not show up as a high-current-density region in the $J_z$ image. This is because this QSL branch is associated with low-current densities (the level is $J_z \sim 1.2$ so that the region disappears with the present gray color-scale used in Fig.~\ref{fig3}a). Note, however, that in general, there is not necessarily a one-to-one association between QSLs and current-density structures, because potential (current-free) magnetic configurations can be associated with QSLs (see also the discussion in Fig. \ref{fig1} above). Here, currents are present in most of the QSLs. The large double-J hooked QSLs are bounded by white regions corresponding to magnetic field lines that reach the lateral boundaries of the simulation box, and blue areas correspond to weakly distorted magnetic field lines. 

The large hooked structure of the QSL in the negative polarity stretches toward lower $y$-value regions but remains within the simulation box. Some QSLs are also identified along the PIL and correspond to the bald patch region, similarly to what was found for the current. In the negative polarity, other QSL footpoints extend toward high $y$-values, but are not associated with reconnecting field lines. They are instead reminiscent of the asymmetric photospheric shear that resulted in the flux rope build-up. This analysis permits us to highlight in Fig. \ref{fig3}c the main stripes of QSLs associated with 3D reconnection with thick dashed pink lines.
The MHD simulation therefore reveals that intense $Q$ layers around the PIL are localized where the intense direct current ribbons are found. Since QSLs are associated with 3D reconnection, this also gives strong evidence that current ribbons can be associated with flare ribbons. The QSL structure discussed here is very similar to the configuration found for a pre-eruptive magnetic field configuration in \citet{Schrijver11} and as depicted in their Figure 21.

\subsection{J-hooked structures and evolution in time}
\label{subsec24}

As shown in Fig. 6 of \citetalias{Aulanier12}, the strong direct currents in the center of the positive and negative polarities are linked with forward J-shaped sigmoidal magnetic field lines (\ie, weakly twisted field lines associated with the erupting flux rope). This double J-shaped current structure found in the present simulation can be related with observations of sigmoids such as reported in \citet{Green11}, and are similar to observations and numerical models discussed in \citet{Savcheva12}. The present simulation shows the evolution in time of 3D magnetic fields formerly associated with a sigmoidal pre-eruptive phase similarly as \citet{Savcheva12}, and as depicted in the Figure 21 of  \citet{Schrijver11}.

Intense $Q$-layers with a double-J structure in the photosphere are also identified within the two magnetic polarities. This configuration is a signature of twisted flux tubes, as was shown analytically by \citet{Demoulin96b} and \citet{Titov07}. 
 Field lines anchored in the regions surrounded by the hooks of the QSLs, and therefore of the double-J currents, correspond to the expanding flux rope. Its tear-drop cross-section is shown in the description of Fig.\ref{fig1}, middle row.
The elongated QSL stripes at $z=0$, shown with the pink lines in Fig.~\ref{fig3}c, form the two principal branches of the X-shaped HFT as seen in Fig. \ref{fig2}. Similarly, the current ribbons depicted in panel a of Fig. \ref{fig3} are linked with the 2D cuts of electric current of Fig. \ref{fig1}: the footpoints of the cusp, which defines the frontier between pre- and post-reconnected magnetic field lines, are situated within these current ribbons.

As discussed in \citetalias{Aulanier12}, the hooks of the J-shaped photospheric currents almost do not evolve in time, while the current ribbons eventually move away from the PIL. This motion can be readily seen in Fig. \ref{fig1} where the legs of the cusp move outward. This motion is again very similar to that of flare ribbons observed during eruptive flares \citep[\textit{e.g.}][]{Wang03}. 

Comparisons of the locations of high-current density build-up and high squashing degree from Figs. \ref{fig1}, \ref{fig2} and \ref{fig3} reveal strong similarities during the whole time-evolution: high $|J|$ locations are associated with regions of high $Q$. Correspondence in shapes can also be pointed out, such as the J-hooked structures seen in the photospheric maps in Fig. \ref{fig3} and the cusps in Fig. \ref{fig1} and QSLs and electric currents evolve similarly in time. The present simulation shows that QSLs are preferential locations for a thin current layer to build and therefore for 3D reconnection to take place, as found in previous simulations \citep{Aulanier05b,Buchner06, Masson09,Wilmot09, Wilmot10} and very recently in laboratory plasmas \citep{Gekelman12}, all in non-eruptive configurations.


\section{Slip-running 3D reconnection}
\label{sec3}

\subsection{Flare loops and flux rope formation via slipping reconnection}

    \begin{figure*}
     \includegraphics[bb=20 270 630 630,width=1.00\textwidth,clip]{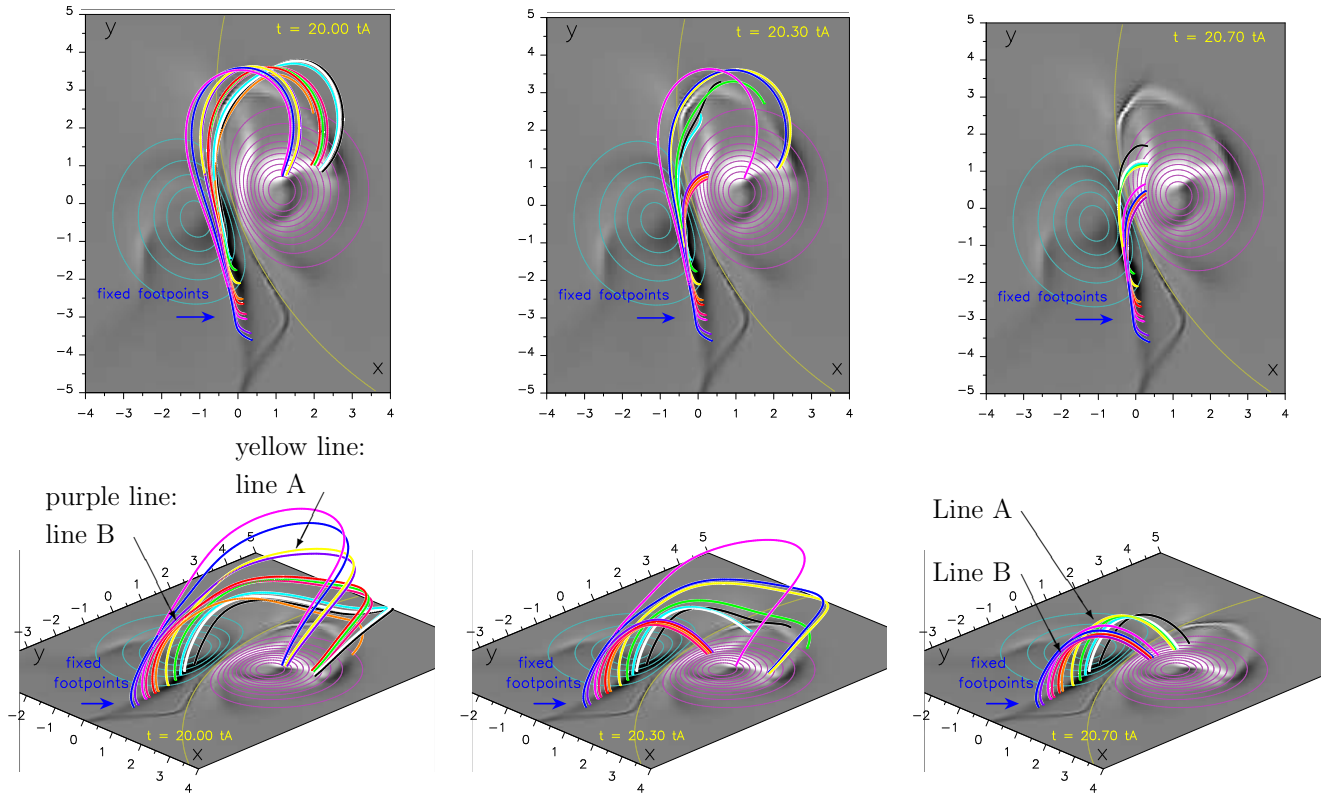}
     \caption{
     Flare loop formation via slip-running reconnection.
   Top and side views show a set of field lines drawn at $t = 20, 20.3, 20.7\ t_A$. The field lines are integrated at each time from the same set of footpoints (anchored at their respective coordinates $F_-$). The panels show different locations of the moving footpoints within the positive magnetic polarity. The gray-scale image and the cyan/pink contourplots represent $J_z(z=0)$ and $B_z(z=0)$ as in Fig. \ref{fig3}a. 
	}
     \label{fig4}
     \end{figure*}

   \begin{figure*}
     \includegraphics[bb=10 370 590 710,width=1.00\textwidth,clip]{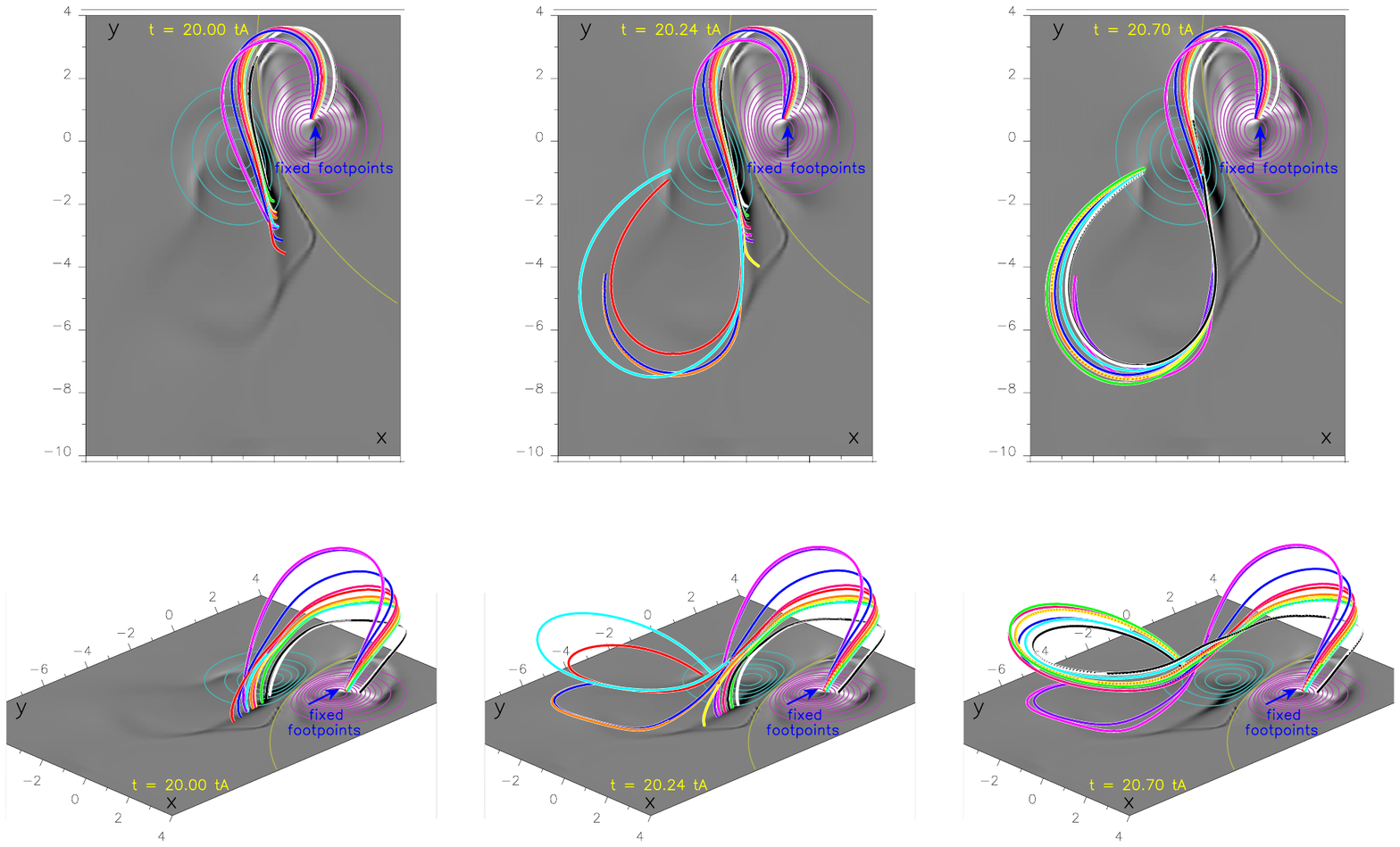}
     \caption{
     Flux rope growth via slip-running reconnection.
   Similar views as in Fig. \ref{fig4} show a set of field lines drawn at $t = 20, 20.24, 20.7\ t_A$. The field lines are integrated from fixed footpoints anchored at $F_+$ in the positive polarity and the panels show the locations of the moving footpoints within the negative magnetic polarity. 
   The views are extended to show the full field line extension.	
   }
     \label{fig5}
     \end{figure*}

We now investigate the features of the reconnection process that leads to the formation of flare loops during the flux rope ejection.
To study the 3D reconnection at work here in detail, we generated outputs in the simulation every $\Delta t=0.02\ t_A$.

Top and side views in Fig. \ref{fig4} show selected field lines at different times with their footpoints anchored in the negative polarity with line-tied conditions at $z=0$. In the following, we write $F_-$ ($F_+$) the coordinates of the footpoint in the negative (positive) polarity. Note that the fixed footpoints are anchored in the strong direct current patch (or ribbon) as described in Sect. \ref{subsec23}. Their conjugated footpoints in the positive polarity evolve significantly, but only during a short time-interval. This evolution occurs when the QSL moves, following the outward motion of the current ribbon (as explained in \ref{subsec23}) and overtakes the footpoint at $F_-$. Figure \ref{fig4} illustrates the field line motion from an initial position in the positive polarity $F_+(t=20\ t_A)$ to a final location $F_+(t=20.7\ t_A)$. After $t=20.7\ t_A$, no significant motion of the field lines can be detected. The moving field lines eventually form strongly sheared flare loops (right column). The field lines do not jump from $F_+(t=20\ t_A)$ to $F_+(t=20.7\ t_A)$ instantaneously, but are seen to ``slip'' in time. An intermediate time at $t=20.3\ t_A$ (middle column) is shown to illustrate this continuous change of connectivity. 

At $t=20\ t_A$, the locations $F_+$ of the footpoints in the positive polarity are all in the strong direct current patch in the middle of the magnetic polarity. At the intermediate time $t=20.3\ t_A$, some of the footpoints remain in this patch, while others have moved along the hook of the $J$-shaped patch. Some are also located in the intense direct current ribbon close to the PIL. At $t=20.7\ t_A$, all the footpoints $F_+$ are located in the direct current ribbon. The footpoints in the positive polarity of the successively reconnected field lines are therefore always located in the intense direct current. They also follow the hook-shaped QSL (pink dashed line of Fig. \ref{fig3}) during their motion. The flare loops eventually formed via the continuous series of reconnection are then anchored in the direct current ribbons along the PIL: this finally confirms the relation between current and flare ribbons.

A different set of field lines is presented in Fig. \ref{fig5}. Now, the footpoints are anchored and fixed in time in the positive polarity at coordinates $F_+$. These correspond to locations within the intense direct current patch in the center of the magnetic positive polarity. The motion of the footpoints in the negative polarity are tracked by field line integration from $F_+$. The colors chosen here for the field lines are unrelated to Fig. \ref{fig4}. The top and side views show the flipping motion at three different times. At $t=20\ t_A$, the footpoints in the negative polarity are all located within the intense direct current ribbon near the PIL. At intermediate times, some of them remain relatively close to their initial position, while other footpoints are located in the hook-shaped and elongated current patch on the other side of the negative polarity. At $t=20.7\ t_A$, most of the footpoints are located in the current patch near the center of the negative magnetic polarity. Similarly to Fig. \ref{fig4}, the field lines move within the current patches and follow the hook-shaped QSL (Fig. \ref{fig3}). We can then point out that, for the flux rope and flare loops formation, QSL and high-$J$ regions with similar structures also play similar roles.
From the side views of Fig. \ref{fig5} (bottom row), the field lines are seen to wind in on themselves and they are also winding around the flux rope axis (as seen also in Fig.~5 in \citetalias{Aulanier12}). They are therefore associated with the expanding and growing flux rope. Eventually, the field lines created by the continuous series of magnetic reconnection form different layers of the flux rope envelope.
The sets of field lines corresponding to the flare loops and the flux rope are shown in separate figures, but they belong to the same reconnecting field lines pairs, as shown in Fig. 5 of \citetalias{Aulanier12}.

The apparent slippage of magnetic field lines is not the same for all field lines. In Fig. \ref{fig4}, at the intermediate time $t=20.3\ t_A$, the position $F_+$ of the red line is located in the direct current ribbon near the PIL, while $F_+$ of the yellow line is still located in the direct current patch in the center of the magnetic polarity. This is in sharp contrast with $t=20.24\ t_A$, when they are both located at similar positions.
Similarly, in Fig. \ref{fig5}, the position $F_-$ of the red line at $t=20.24\ t_A$ is located in the direct current patch near the center of the negative magnetic polarity, while the yellow line footpoint is located in the direct current ribbon near the PIL. Thus, the moving footpoints are not shifted by the same distance during the flipping motion. Furthermore, tracking their motion from the initial to their final position reveals that they do not have the same speed. Since the field lines are slipping from the initial to the final positions within less than one Alfv\'en time, this motion can be coined a ``slip-running'' reconnection motion (as defined in \citealp{Aulanier06}).
 In the following, we focus on the slip-running speed characteristics by studying the motion of two selected field lines associated with flare loops: the yellow and purple ones indicated in Fig. \ref{fig4}.  Hereafter, they are referred to as line A and line B, respectively.

\subsection{Super-alfv\'enic slippage of selected field lines}

      \begin{figure*}
     \sidecaption
     \includegraphics[bb=35 270 570 600,width=1.00\textwidth,clip]{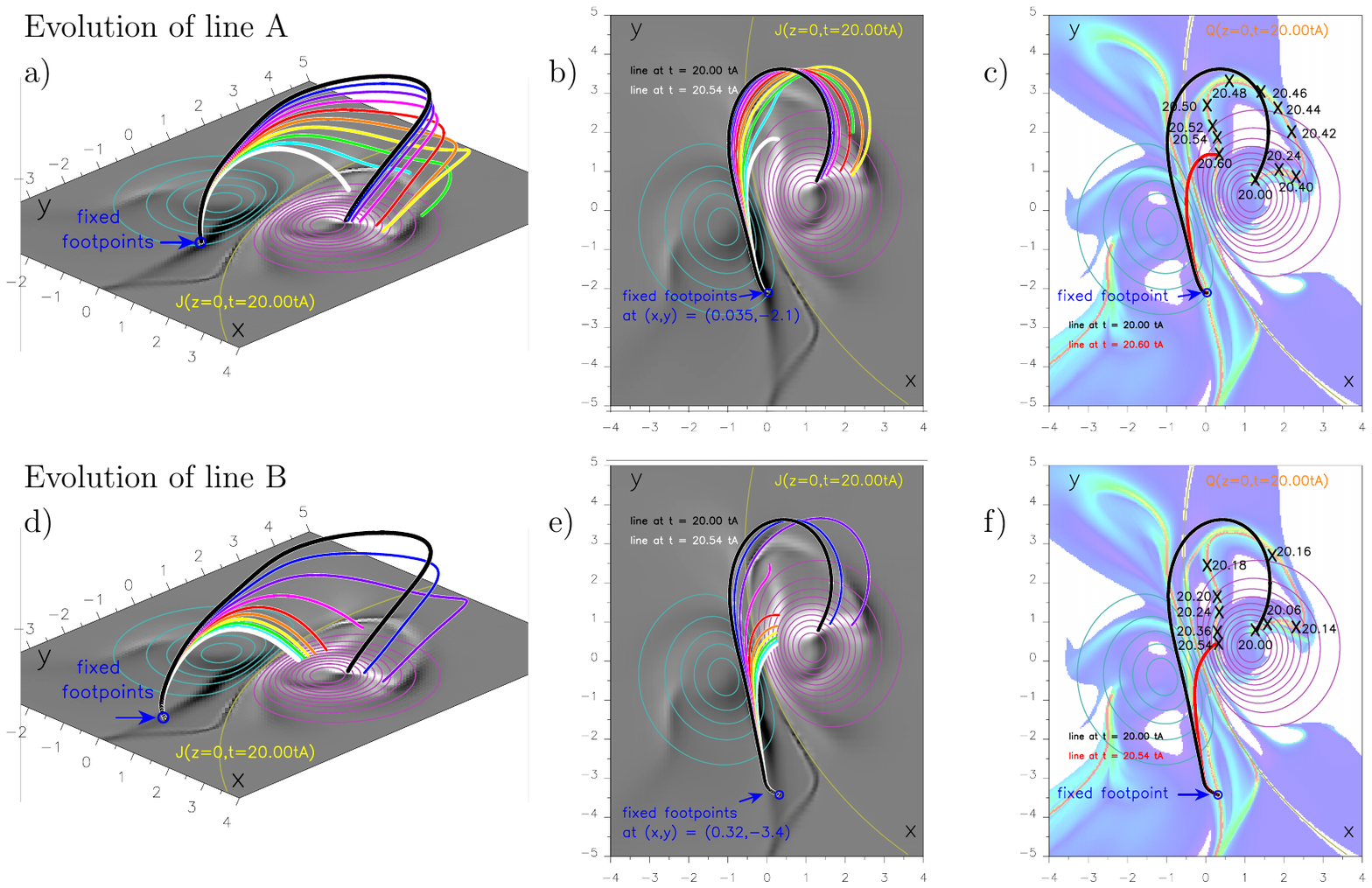}
     \caption{
    Time evolution during the slip-running motion of field lines A and B.
Side (\textbf{a}) and top (\textbf{b}) views of the line A defined from Fig. \ref{fig4} with fixed $F_-(x=0.035,y=-2.1)$. The set of colors varies from black to white in a rainbow fashion, and corresponds to line A at every $\Delta t = 0.06\ t_A$ from $t = 20\ t_A$ (in black) to $t=20.54\ t_A$ (in white). The gray-scale image and the cyan/pink contourplots correspond to $J_z(z=0)$ and $B_z(z=0)$ at $t=20\ t_A$ with the same color coding as in Fig. \ref{fig3}. (\textbf{c}) A top view  of the QSLs at $t=20\ t_A$ with the same color-coding as in Fig. \ref{fig3}b (but shaded) is presented with the locations of the moving $F_+$ at different times. The field line at $t=20\ t_A$ and at $t=20.6\ t_A$ is represented in black and red.
(\textbf{d,e,f}): same views for line B from Fig. \ref{fig4} with its fixed footpoint $F_-(x=0.32,y=-3.4)$. The color-coding is the same as for a.b.c.
	}
     \label{fig6}
     \end{figure*}

Figure \ref{fig6} shows field lines A and B with fixed footpoints in the negative magnetic polarity (as shown in Fig. \ref{fig4}). 
Figure~\ref{fig6}a shows line A at different but equidistant times from $t=20\ t_A$ (in black) to $t=20.54\ t_A$ (in white) but all drawn together on a single panel. We used a rainbow color gradation to distinguish the different times and to follow the motion of the footpoint in the positive polarity. Figure~\ref{fig6}b shows the same outputs from a top view. Figure~\ref{fig6}c represents a shaded QSL map (\ie, $\log Q (z=0)$ at $t=20\ t_A$) and the different locations of $F_+$ and their corresponding times. The same views are presented in the bottom row for line B.

The two field lines A and B show similar characteristics during their slippage. The inverse J-hooked shape of the initial position (in black) is present in both cases (Figs.~\ref{fig6}b and e), and the final positions correspond to sheared flare loops (in red, Figs.~\ref{fig6}c and f).
The two footpoints in the positive polarity move inside the thin QSL footprint, following the hook shape of the QSL. Their motion is faster around the hook of the QSL, since the distance between successive footpoint locations is larger there (Fig. \ref{fig6}).

The slip-running motion of lines A and B is related to the motion of the QSL in the negative polarity away from the PIL: as the QSL overtakes the field line footpoints, these field lines slip-run in time. When looking at the QSL map of Fig. \ref{fig3}b in the region surrounding the two footpoint locations $F_-$ for the two lines,  we see that faster slip-running motion corresponds to $F_-$ located in the intense $Q$ layer. Note also that the distance between successive footpoint locations in the QSL hook is larger for line B than for line~A. Therefore, line B slip-runs faster than line A. More generally, analyzing other field lines shows that the slippage depends on the field line and its $F_-$ coordinates.

\begin{figure*}
    \sidecaption
      \includegraphics[bb=35 268 555 600,width=12cm,clip]{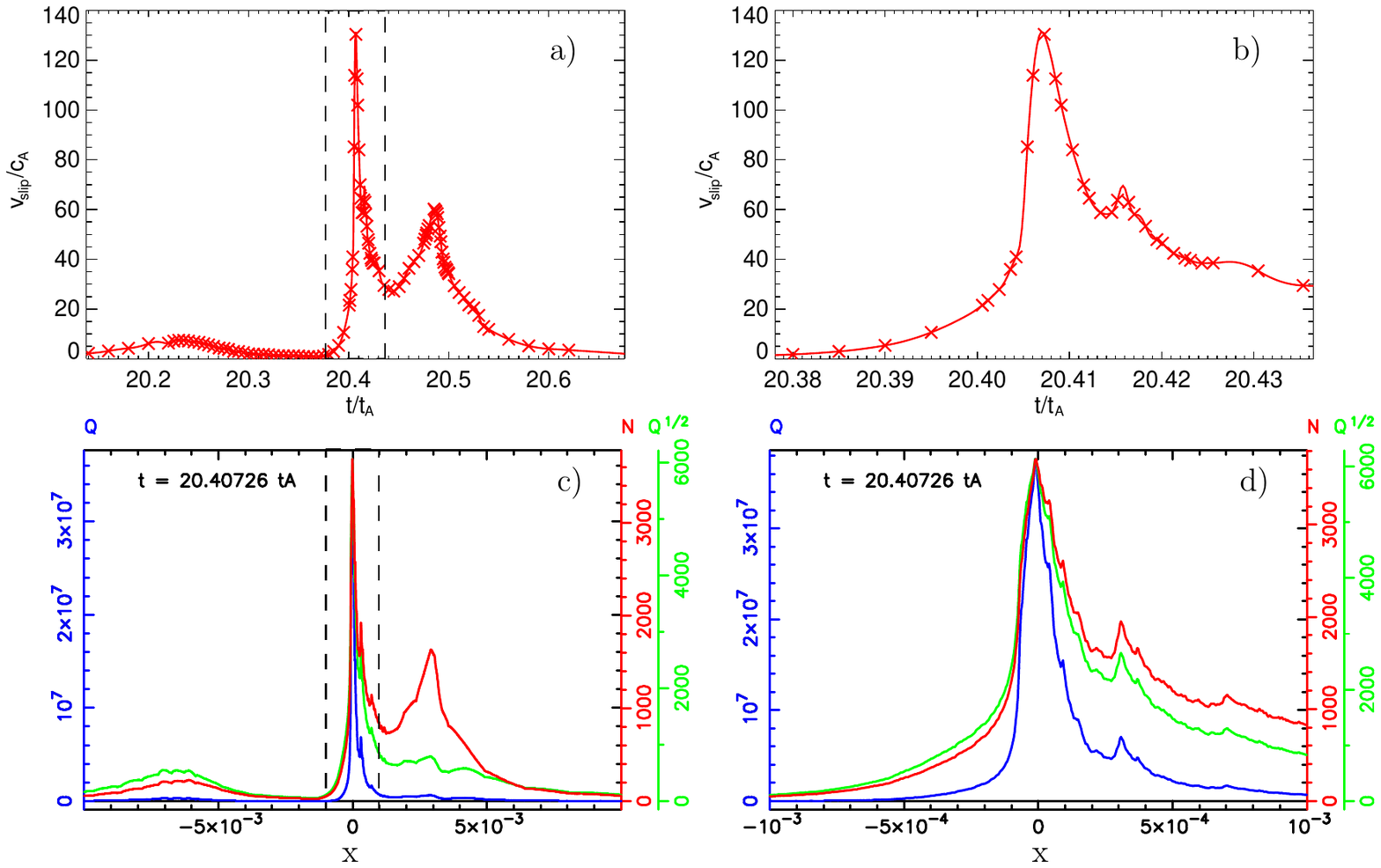}
     \caption{
     Velocity and QSL analysis for field line A of Fig. \ref{fig4}.
     \textbf{a)} Time evolution of the positive footpoint velocity \vslip, centered on the central peak (\vpeak) reached at $t=20.407\ t_A$. \textbf{b)} Zoom, within the dashed lines of panel a, showing a secondary maximum close to the main peak. \textbf{c)} 1D $x$-cut of $N,Q$ and $Q^{1/2}$ (in red, blue, and green) centered on $F_-$ of line A and at $t=20.407\ t_A$. \textbf{d)} Zoom on the main peak of $N$, $Q$, and \Qroot\ showing similar features as \vslip. 
     }
     \label{fig7}
    \end{figure*}

\begin{figure*}
      \sidecaption
      \includegraphics[bb=35 268 555 600,width=12cm,clip]{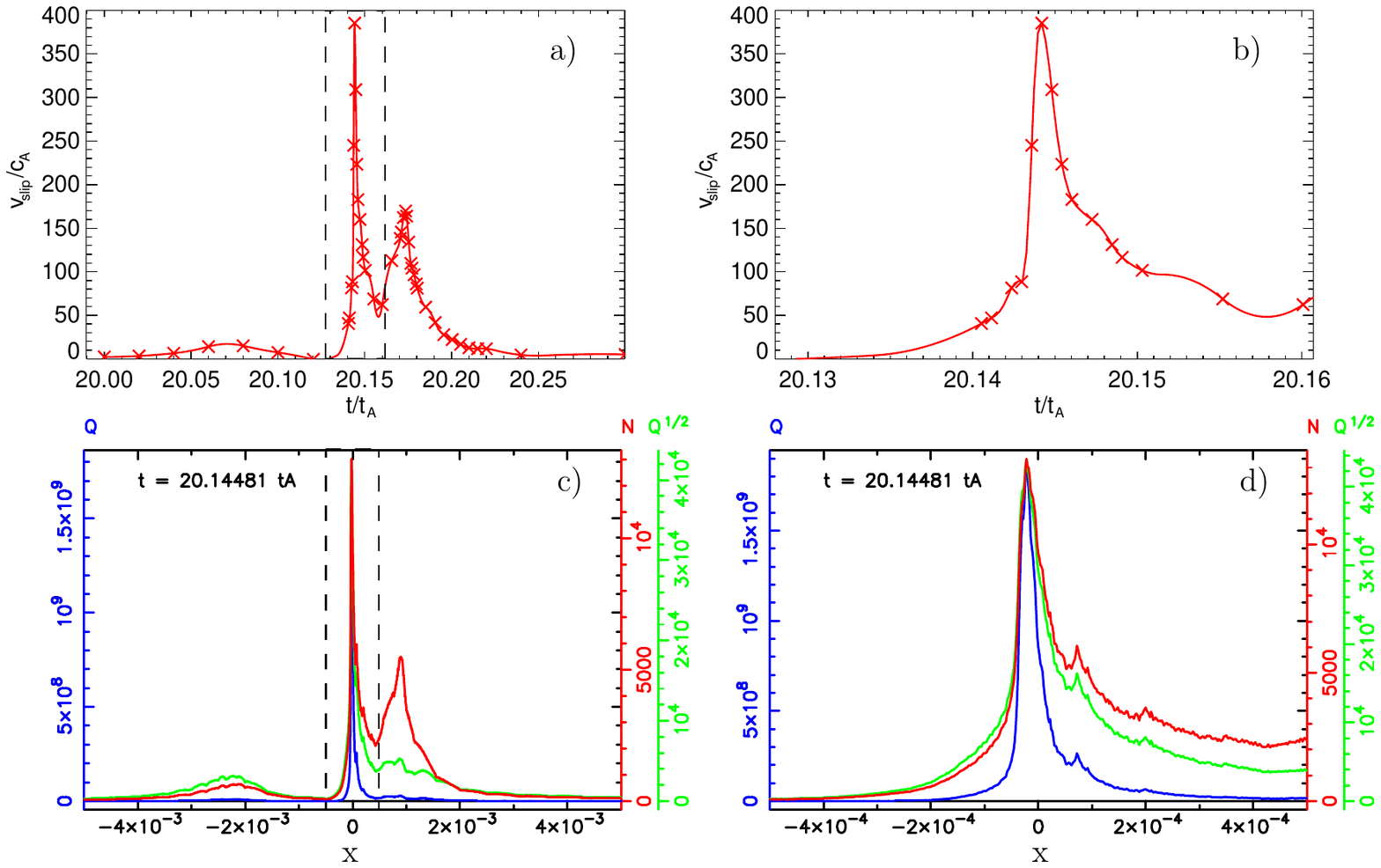}
     \caption{
          Velocity and QSL analysis for the field line B of Fig. \ref{fig4}, with the central peak (\vpeak) reached at $t=20.145\ t_A$. Each panel is drawn according to the same convention as in Fig. \ref{fig7}.
}
     \label{fig8}
    \end{figure*}

Knowing the $F_+$ positions and the related times, we computed the slip-running speed \vslip\ profile for the two field lines. These profiles are given in the top rows of Fig. \ref{fig7} (field line A) and \ref{fig8} (field line B). The outputs, marked with $\times$ signs, were calculated every $\Delta t=2\times 10^{-2}\ t_A$ to  every $\Delta t=5\times 10^{-3}\ t_A$ (and to $\Delta t=7\times 10^{-4}\ t_A$ for line B) to resolve the different narrow peaks in the speed profiles. We also used a third-order spline interpolation between the measured positions $F_+$ to obtain the speed evolution in time. All the secondary peaks are clearly resolved, and the central peak for line B is marginally well resolved. Since the resolution highly depends on $\Delta t$, obtaining an extremely refined central peak is cumbersome (simulation-wise), but the following Sect. \ref{sec4} shows that the results found with the present $\Delta t$ are quantitatively satisfying.
The analysis of Figs. \ref{fig7} and \ref{fig8} leads to two main conclusions.

Firstly, the slip-running motion varies in time, as can be seen in Figs. \ref{fig7}a and \ref{fig8}a. This evolution is related with the displacement of the QSL as discussed above, since the fixed footpoint in the negative polarity $F_-$ is swiped by the moving QSL. We verified that the central peak of \vslip\ (hereafter noted \vpeak) is reached at the same time at which $F_-$ was anchored in the high-$Q$ region (central red layer in Fig. \ref{fig3}b). This indicates that QSL geometries play a strong role in determining \vpeak\ and the time evolution of \vslip.

Secondly, the time evolution of \vslip\ shows that $v_{\mathrm{peak}}(\mathrm{B}) \sim 390\ $\cA $\ > v_{\mathrm{peak}}(\mathrm{A})\sim 130\ $\cA, \ie, line B slips faster than line A. Still, the two speed profiles present similar features: first a low peak $<20$ \cA\ (at $t\sim 20.25\ t_A$ for line~A, and $t\sim 20.055\ t_A$ for line B, see Figs. \ref{fig7}a and \ref{fig8}a), and a central peak followed by a second high peak \vslip\ $\approx$ \vpeak/2. The zoom on the central peak for line A (Fig. \ref{fig7}b) in fact shows another peak, very close to \vpeak, that is only suggested in the velocity profile for line~B (Fig.~\ref{fig8}b) due to the resolution of the speed profile (even with the lower $\Delta t$ used here). Note that such high slip-running speed values have never been calculated in simulations before. Considering the short time intervals needed to quantify \vslip, the slip-running reconnection regime is difficult to detect in simulations running with a time interval on the order of one Alfv\'en time.
In the following, we quantify the similarities found between \vslip\ and the parameters that define the QSLs.

\subsection{Similarities in the \vslip\ profiles and 1D cuts of the QSL}
\label{subsec33}

The footpoints $F_+$ for the lines A and B are moving within the same QSL in the positive polarity (right column, Fig.~\ref{fig6}).
However, the QSL map presented in Fig. \ref{fig6} cannot give a precise quantification of the squashing degree $Q$ because of its low spatial resolution. To quantify $Q$ and the mapping norm $N$, high spatial resolution of 1D cuts of the QSL around $F_-$ were made and are presented in the bottom rows of Figs. \ref{fig7} (line A) and~\ref{fig8} (line B). Note that in contrast to the \vslip\ profile, the convergence for the field line integration has been tested for the 1D cuts of the QSLs to resolve the fine peaks of $N(x)$ and $Q(x)$.

Figures \ref{fig7}c and \ref{fig8}c represent the spatial distribution of the mapping norm $N$ (in red), the squashing degree $Q$ (in blue), and $Q^{1/2}$ (in green, drawn here because it is directly proportional to $N$, see subsection \textit{2.2}). The graphs are centered on $x_{\mathrm{peak}}$, with $x_{\mathrm{peak}}=0$ when \Npeak\ (and $Q_{\mathrm{peak}}$) is reached. Interestingly, $ x(F_-)$ is at $x_{\mathrm{peak}}$ for both lines at the time at which \vpeak\ is reached, demonstrating that \vslip\ peaks for the highest values of $N$ and $Q$.

The spatial range of these panels is $\pm 10^{-2}$ ($\pm 5\times 10^{-3}$) around $x_{\mathrm{peak}}$ for line A (line B). This frame is chosen to show the strong similarities existing between the 1D spatial cuts and the \vslip\ profiles of Figs. \ref{fig7}a and \ref{fig8}a.
Comparing the two graphs shows that the $N$ spatial distribution is closely similar to that of \vslip.

Such similarities can also be found with a finer resolution. The panels \ref{fig7}d and \ref{fig8}d present a zoom of the highest peak for $N,Q,Q^{1/2}$ with a window $\pm 10^{-3}$ around $x_{\mathrm{peak}}$ for line A and $\pm 5\times10^{-4}$ for line B. Similar secondary peaks appear for the $N$ and \vslip\ profiles for line A (Fig. \ref{fig7}b,d), and we would expect similar features for line B (Fig. \ref{fig8}b,d) with a better temporal resolution of the profile surrounding \vpeak.

All similarities with the \vslip\ profile are stronger for $N$ for the full $x$ interval shown here (Figs. \ref{fig7}c, \ref{fig8}c). The second-highest peak in the slip-running motion profile appears more clearly in the $N$ spatial distribution than for $Q$ and \Qroot.  What happens is that even though \Qroot\ is directly related to $N$, their profiles are actually similar to each other only on the left side of the graphs (see Figs. \ref{fig7}c and \ref{fig8}c), \ie\ for $x \lesssim x_{\mathrm{peak}}$. The reason is that the profile of $B_z^{\mathrm{ratio}}$ (not presented here) is almost constant only for $x\lesssim x_{\mathrm{peak}}$ and not for $x > x_{\mathrm{peak}}$. So $Q^{1/2}$ is directly proportional to $N$ ($Q^{1/2} = N/B_{\mathrm{const}}^{1/2}$) only for $x \lesssim x_{\mathrm{peak}}$. This linear relation disappears for $x>x_{\mathrm{peak}}$, which explains the difference in the two profiles. This suggests that the mapping norm $N$ is a better proxy to relate the slip-running motion to the QSLs. Section \ref{sec4} provides a more precise analysis.

Finally, the 1D cuts represented in Figs. \ref{fig7}c and \ref{fig8}c correspond to the same QSL footpoint but at a different $y$ position. This first explains the qualitatively similar features in the $N, Q, and Q^{1/2}$ profiles found in both the $y=-2.1$ cut (Fig. \ref{fig7}c) and the $y=-3.4$ cut (Fig. \ref{fig8}c). However, the peak at $y=-3.4$ is much narrower than at $y=-2.1$. Since the local maximum value \Npeak\ is simply linked with the local thickness of the QSL (see Eq. (21) in \citealp{Demoulin96b}), $N_{\mathrm{peak}}$ and $Q_{\mathrm{peak}}$ values strongly increase as $y$ diminishes. This explains why $v_{\mathrm{peak}}(\mathrm{B}) > v_{\mathrm{peak}}(\mathrm{A})$.

The present study quantifies the qualitative changes of the QSL that were seen in Fig. \ref{fig3}. On the one hand, $v_{\mathrm{peak}}$ is reached when the fixed footpoint $F_-$ is swiped by the \Npeak (or $Q_{\mathrm{peak}}$) layer. On the other hand, the \vslip\ profile has very similar features with that of the norm $N$. Section \ref{sec4} investigates the correlation between \vslip\ and $N$ and $Q$ more quantitatively.
        \begin{figure*}
         \sidecaption
      \includegraphics[bb=40 425 600 620,width=12cm,clip]{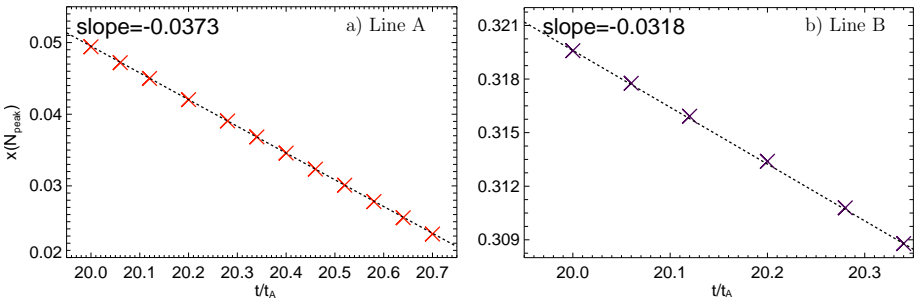}
     \caption{Horizontal displacement of the QSL in time along $x$ around $F_-$ for line A and B.
     \textbf{a)} Time evolution of coordinate $x$(\Npeak) for $y(F_-)=-2.1$, showing the displacement of the QSL around $F_-$ for field line A. \textbf{b)} Same for $y(F_-)=-3.4$ for field line B.}
     \label{fig9}
    \end{figure*}

\section{\vslip\ and $N$ correlation and their time evolution}
\label{sec4}

\subsection{Correlation between $v_{\mathrm{slip}}$ and $N$}

Figures \ref{fig7} and \ref{fig8} show that the time evolution of \vslip\ within a short time interval ($\Delta t < 1\ t_A$) is similar to the spatial changes of the norm $N$ and the squashing degree \Qroot, although the similarities are clearly much stronger for $N$.
We analyze this relationship below in more detail.
This is done by considering the displacement of the QSLs: the latter eventually move outward from the PIL, similarly to the flare ribbons (as discussed in Sect. \ref{sec2}). This photospheric motion can be seen along the $(y=-0.3,z=0)$ cuts shown in the middle row of Fig. \ref{fig1}.

First, to calculate the speed related to the displacement of the QSLs, the reference point \Npeak\ was chosen. Then, different $x$-cuts of the QSL (such as panel \ref{fig7}c) were made at different times, which allowed us to track the displacement of the peak position $x($\Npeak) along $x$. Figure \ref{fig9} shows this displacement for line A (the $x$-cuts are made at $y=-2.1$) and line B (the $x$-cuts are made at $y=-3.4$). In both cases, the displacement of the QSL is a linear function of time, \ie, the speed \vQSL$=|\mathrm{slope}|$ is constant within the time intervals of Figs. \ref{fig7}a and \ref{fig8}a. We obtain \vQSL$=3.73\% $~\cA\ for line A and $3.18\%$ \cA\ for line B. This provides a direct relation between the spatial distribution of $N$ and \Qroot\ and the time distribution of \vslip, both shown in Figs. \ref{fig7} and \ref{fig8} (because it connects the abscissa $x$ and $t/t_A$ of the panels).

     \begin{figure*}
     \sidecaption
      \includegraphics[bb=50 250 590 610,width=12cm,clip]{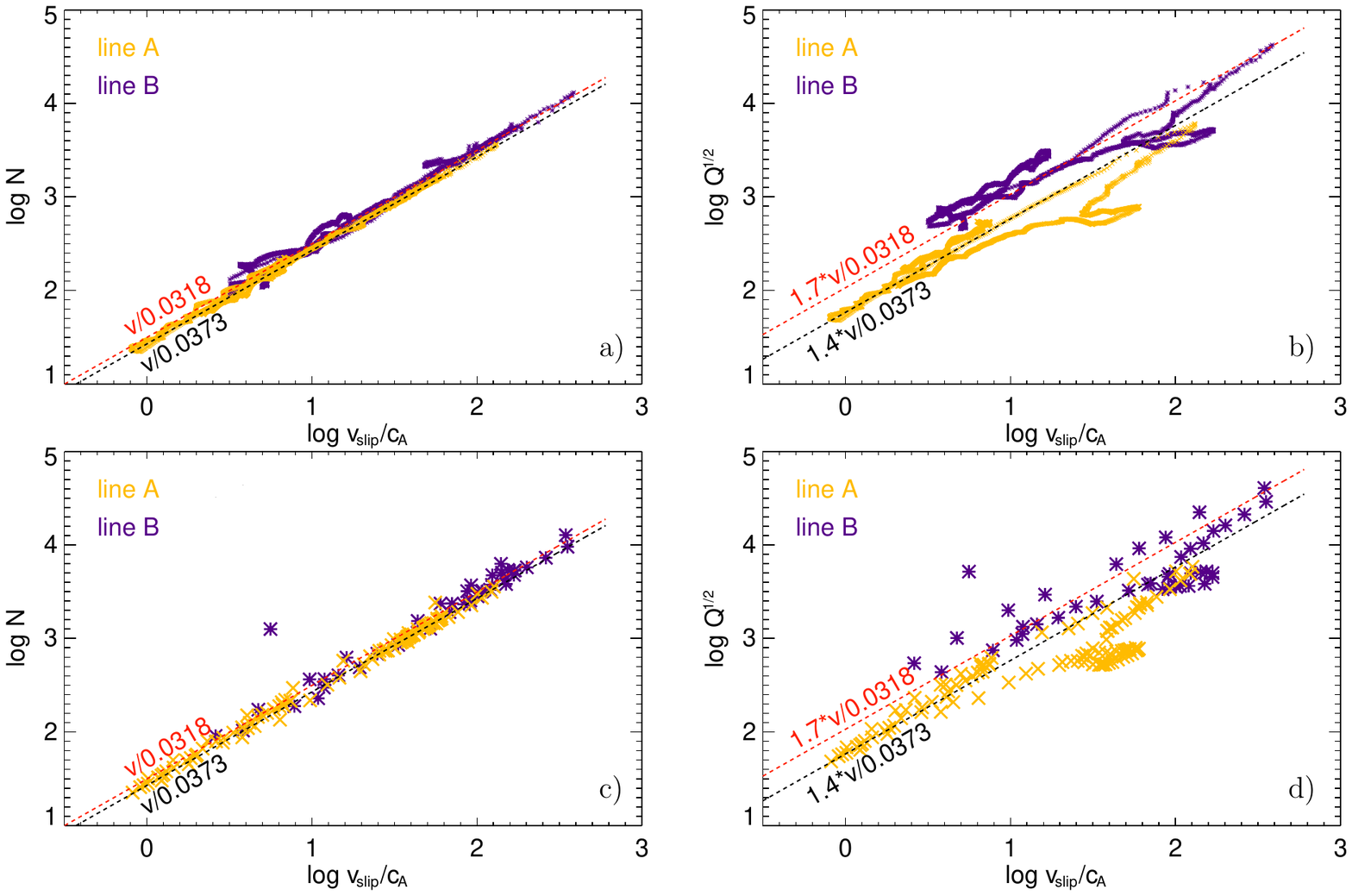}
     \caption{Linear correlation between $\log N$, $\log$ \Qroot\ and $\log$ \vslip.
     \textbf{a)} $\log N$ as a function of $\log (v_{\mathrm{slip}}/$\cA) for Line A (yellow) and Line B (purple) with \vslip\ obtained from third-order spline interpolation. \textbf{b)} $\log$ \Qroot\ in function of $\log (v_{\mathrm{slip}}/$\cA\ with the same color-code and for the same interpolation scheme. \textbf{c),d)} Same graphs, but with \vslip\ measured with a linear interpolation method. The function $f($\vslip)=$\log$(\vslip)$-\log$(\vQSL)\ is added to all the graphs, with \vQSL\ = 0.0373 for line~A (black) and \vQSL\ = 0.0318 for line~B (red), see Fig. \ref{fig9}.
     }
     \label{fig10}
    \end{figure*}

\begin{table}[t]
\caption{Pearson and Spearman correlation coefficients $c_\mathrm{P}$ and $c_\mathrm{S}$ for $N$ vs \vslip\ and \Qroot\ vs \vslip.}
\begin{tabular}{r p{0.1\textwidth}  p{0.1\textwidth}}
\hline
\hline
   Spline interpolation & $\ \ \ \ \ \ c_\mathrm{P}$  & $\ \ c_\mathrm{S}$ \\\hline
\multirow{2}{0.1\textwidth}{Line A} $N$ vs \vslip   & $\ \ \ \ 0.998$ & 0.996 \\
$Q^{1/2}$ vs \vslip  & $\ \ \ \ 0.932$ & 0.969 \\
\hline
\multirow{2}{0.1\textwidth}{Line B} $N$ vs \vslip & $\ \ \ \ 0.987$  & 0.935 \\
$Q^{1/2}$ vs \vslip  & $\ \ \ \ 0.934$ & 0.923 \\
\hline
\hline
          Linear interpolation  & $\ \ \ \ \ \ c_\mathrm{P}$ & $\ \ c_\mathrm{S}$ \\
\hline
\multirow{2}{0.1\textwidth}{Line A}  $N$ vs \vslip & $\ \ \ \ 0.997$ & 0.997 \\
 $Q^{1/2}$ vs \vslip & $\ \ \ \ 0.913$ & 0.922 \\
\hline
\multirow{2}{0.1\textwidth}{Line B}  $N$ vs \vslip & $\ \ \ \ 0.964$ & 0.976 \\
 $Q^{1/2}$ vs \vslip & $\ \ \ \ 0.842$ & 0.805 \\
\hline
\end{tabular}
\label{tab1}
\end{table}

\begin{table}[t]
\caption{Slope $\alpha$ and coefficient of determination $R^2$ of the linear fitting functions for Fig. \ref{fig10}.}
\begin{tabular}{r p{0.1\textwidth}  p{0.1\textwidth}}
\hline
\hline
   Spline interpolation & $\ \ \ \ \ \ \ \ \alpha$  & $\ R^2$ \\
   \hline
\multirow{2}{0.1\textwidth}{Line A}  $N$ vs \vslip   &  $\ \ \ \ \ \ 0.99$ &  $1.00$\\
$Q^{1/2}$ vs \vslip  &  $\ \ \ \ \ \ 0.69$  &  $0.87$\\
\hline
\multirow{2}{0.1\textwidth}{Line B} $N$ vs \vslip &  $\ \ \ \ \ \ 1.01$  & $0.97$ \\
$Q^{1/2}$ vs \vslip  & $\ \ \ \ \ \ 0.71$ &  $0.87$\\
\hline
\hline
          Linear interpolation  & $\ \ \ \ \ \ \ \ \alpha$ & $\ R^2$ \\
\hline
\multirow{2}{0.1\textwidth}{Line A}  $N$ vs \vslip & $\ \ \ \ \ \ 0.98$ & $1.00$ \\
 $Q^{1/2}$ vs \vslip & $\ \ \ \ \ \ 0.71$ &  $0.83$\\
\hline
\multirow{2}{0.1\textwidth}{Line B}  $N$ vs \vslip & $\ \ \ \ \ \ 0.95$ &  $0.93$\\
 $Q^{1/2}$ vs \vslip & $\ \ \ \ \ \ 0.67$ & $0.71$ \\
\hline
          Linear interpolation & \multirow{2}{*}{$\ \ \ \ \ \ \ \ \alpha$} & \multirow{2}{*}{$\ R^2$} \\
          without outside data &  & \\
\hline
\multirow{2}{0.1\textwidth}{Line B}   $N$ vs \vslip & $\ \ \ \ \ \ 1.02$ &  $0.98$\\
 $Q^{1/2}$ vs \vslip & $\ \ \ \ \ \ 0.73$ &  $0.78$\\
\hline
\end{tabular}
\label{tab2}
\end{table}

Using this relation, we present in Table \ref{tab1} the Pearson and Spearman coefficients $c_{\rm{P}}$ and $c_{\rm{S}}$ for the correlation between $N$ and \vslip, and for \Qroot\ and \vslip\ for line A and line B. Two cases are presented corresponding to two interpolations of the footpoint positions before computing \vslip\ : a third-order spline interpolation between the measured positions $F_+$ and a simple linear interpolation method.
With both methods, the correlation coefficients are best for $N$ vs \vslip, showing that \Qroot\ is less correlated to \vslip\ than $N$. These results confirm the qualitative results obtained from the analysis of the temporal profile of \vslip\ and spatial profile of $N$, presented in Figs. \ref{fig7} and \ref{fig8}. Note that $c_{\rm{P}}$ and $c_{\rm{S}}$ are better for line A than line B. This could be due to the better time resolution of \vslip\ for line A as can be seen by comparing Fig.~\ref{fig7}b and Fig. \ref{fig8}b.

Since the time profile of \vslip\ and $x$-distribution of $N$ are very similar and the correlation coefficients are very good, we tested whether $N$ is a linear function of \vslip. As \Qroot\ is directly related to $N$, we also tested the same hypothesis for \Qroot\ and \vslip.
To show this, Figures \ref{fig10}a and c present $\log N$ versus $\log v_{\mathrm{slip}}$ with \vslip\ obtained from the third-order spline interpolation (panel a) and the linear interpolation (panel c). Panels b and d represent $\log Q^{1/2}$ versus $\log v_{\mathrm{slip}}$ for the same interpolation methods.
We also plot in all graphs the function $f($\vslip)=$\log$(\vslip)-$\log$(\vQSL).

For both lines and the two interpolation cases, the linear relation is readily seen for $N$ vs \vslip\ with a very low dispersion, and the graphs show a linear trend similar to $\log$(\vslip)-$\log$(\vQSL). Note, however, that the linear interpolation of the field-line footpoints implies that one point for line B is outside of the linear correlation. This is because of the low time resolution, which implies errors when determining \vslip. However, such an error nearly disappears with the third order spline interpolation, which smoothes the time distribution of \vslip. 
For \Qroot\ vs \vslip, a linear trend exists similarly to the previous graph, but the dispersion is much higher than for $N$.

We then applied a linear fitting function to the graphs. The slope $\alpha$ and the coefficient of determination $R^2$ are reported for lines A and B in Table \ref{tab2}. We also report the same coefficients for the linear interpolation of \vslip\ for line B where the outside data point, corresponding to the low time resolution calculation, has been removed.
We find that the slopes are very close to $\alpha = 1$ for all $N$ vs \vslip\ graphs with a very good determination coefficient $R^2 \geq 0.97$. However, the linear trend is not as good for \Qroot\ since both $\alpha$ and $R^2$ are not close to 1, showing that the linear correlation is only valid for $N$.

As discussed in Sect. \ref{subsec33}, \Bratio\ is almost constant for $x<x_{\mathrm{peak}}$. There is then a proportional relation between $N$ and \Qroot\ for this space interval. We verified this result by removing all $x>x_{\mathrm{peak}}$, which indeed shows a strict linear correlation similar to $N$ and \vslip\ (not presented here). The dispersion seen in Figs. \ref{fig10}b and  \ref{fig10}d  therefore results from the different values taken by \Bratio.
This analysis is deeper than that in Sect. \ref{subsec23} and confirms that the mapping norm $N$ is a better proxy than the squashing degree $Q$ for linking the QSLs with the slipping motion speed \vslip.

    \begin{figure}
      \includegraphics[bb=40 20 1100 750,width=0.5\textwidth,clip]{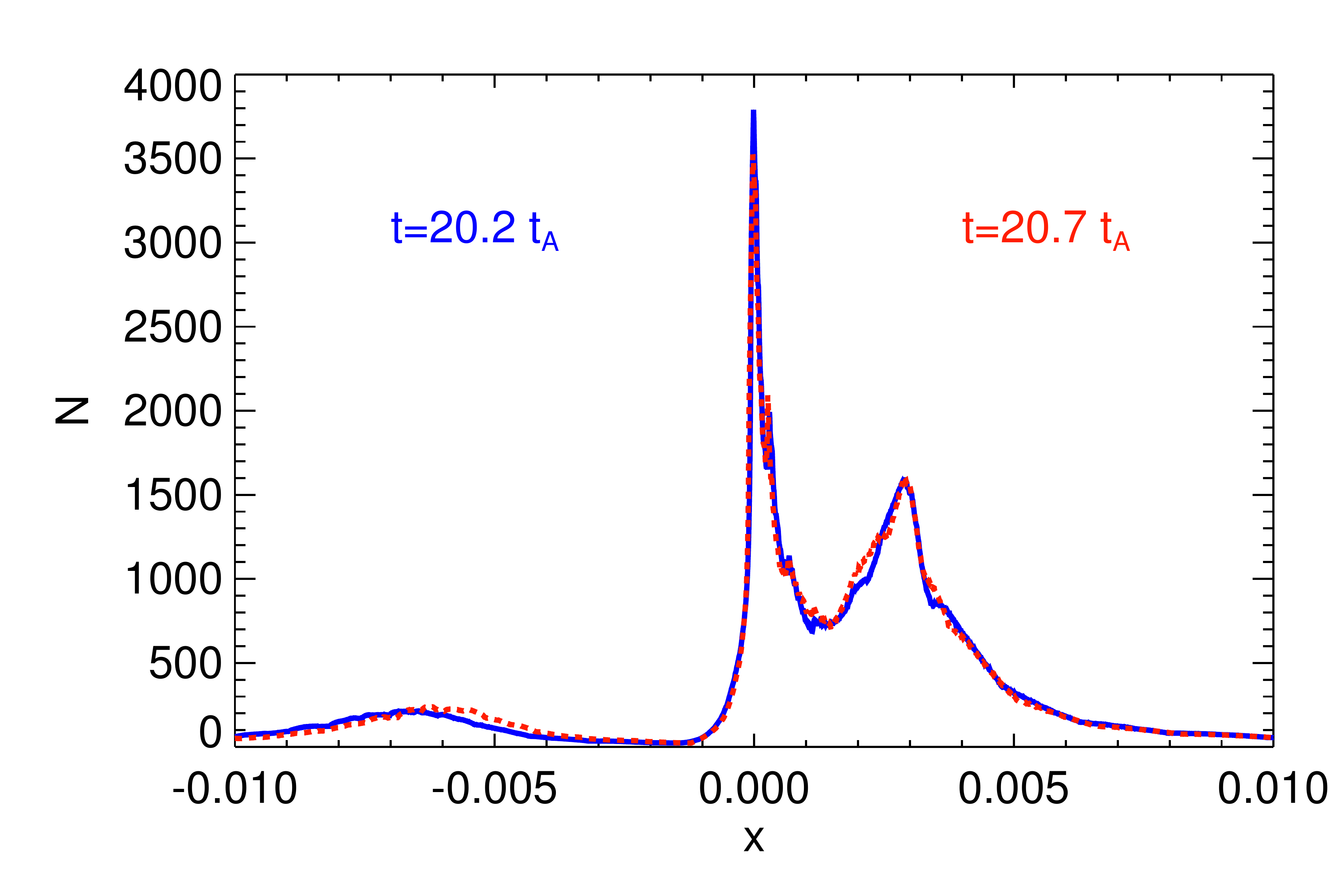}
     \caption{Spatial distribution of $N(x)$ at different times and centered on the peak value. The solid blue line represents $N$ centered on \Npeak\ at $t=20.2\ t_A$ for a cross-section at the footpoint of line A in the negative polarity. The dashed red line represents $N$ for $t=20.7\ t_A$. $N(x)$ almost does not evolve during $0.5\ t_A$.
     }
     \label{fig11}
    \end{figure}

\subsection{Origin of \vslip$=\alpha N$}

What is the meaning of the strict correlation we find between $N$ and \vslip?
To obtain Figs. \ref{fig7}a and \ref{fig8}a, we considered two field lines anchored in the negative polarity at a fixed footpoint $F_-$ that are overtaken by a moving QSL. $F_-$ crosses a portion $\delta l(t)$ of the whole QSL width during a short time interval $\delta t$. As was shown in Fig. \ref{fig9}, the outward velocity of the QSL is $v_{\mathrm{QSL}}(t) =\delta l(t)/\delta t$. It is constant in the time interval of the QSL crossing.
Meanwhile, during the short time interval $\delta t$, the footpoint in the positive polarity $F_+$ moves \textit{along} the conjugate QSL footprint for a fraction $\delta L(t)$ of the whole QSL length. Then, \vslip$(t)=\delta L(t)/\delta t$.
From the definition of the mapping norm $N$, we take only into account the steepest spatial gradients and $N$ is nearly $N(t) \approx \delta L/\delta l$ for $\delta L \gg \delta l$ \citep[see Eq. (3) in][]{Demoulin96a}. 
Then, at time $t$, the two definitions for \vslip\ and $N$ lead to
\begin{equation}
v_{\mathrm{slip}} (t)= \frac{\delta L}{\delta t} = \frac{\delta l}{\delta t}  \frac{\delta L}{\delta l}\approx v_{\mathrm{QSL}} N(t)  .
\label{vslipequation}
\end{equation}

This analytical relation should only be valid at time $t$, since $N(x)$ can change as time passes by. This relation has been verified in Figs. \ref{fig10}a and \ref{fig10}c where the \vslip\ values are related to different times while $N(t = 20.41\ t_A)$ for line A and $N(t=20.15\ t_A)$ are calculated at fixed times. The clear linear correlation found above implies that the spatial distribution of $N$ does not change significantly during the crossing of the full QSL width.
To verify this, we have plotted in Fig. \ref{fig11} the spatial distribution of $N$ for $t=20.2\ t_A$ and $t=20.7\ t_A$ with $x$-cuts made around $F_-$ for line A. The chosen times correspond to the beginning and the end of the slipping motion. Both profiles have been centered on \Npeak\ and have the same spatial window. Indeed, the profiles are very similar, demonstrating that the spatial distribution of $N$ clearly does not change much during the slipping motion.

Note that line A and line B are not specific cases: studying the other lines of Fig. \ref{fig4} shows the same correlation between the slipping motion speed and the mapping field line parameter $N$.

    \begin{figure}
     \includegraphics[bb=60 30 1100 700,width=0.50\textwidth,clip]{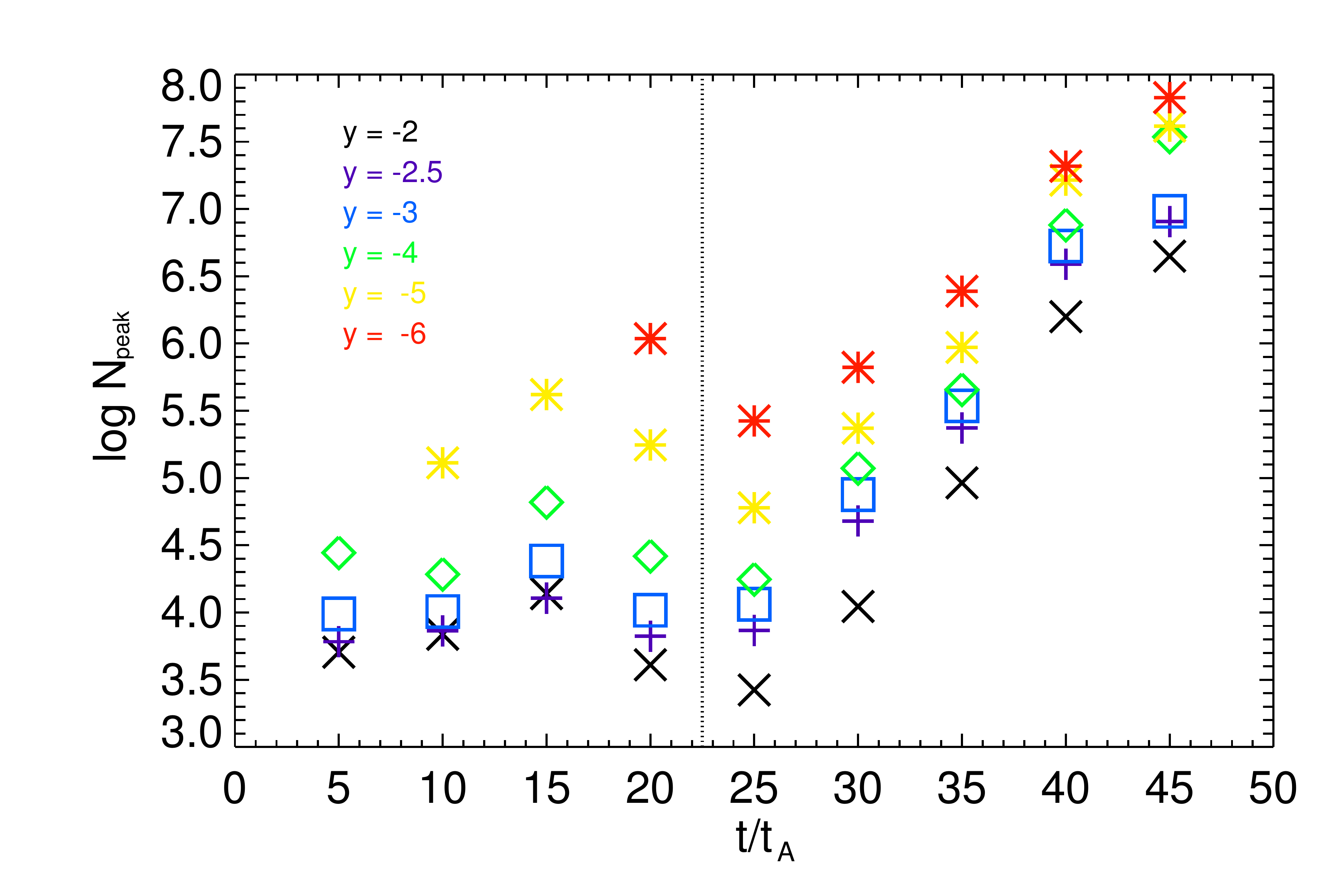}
	\caption{Time evolution of \Npeak\ during the flux rope ejection. The \Npeak\ values are reported by making $x$-cuts at different $y$ (different colors/markers).
         }
 \label{fig12}
  \end{figure}

    \begin{figure*}
     \sidecaption
      \includegraphics[bb=0 250 600 750,width=12cm,clip]{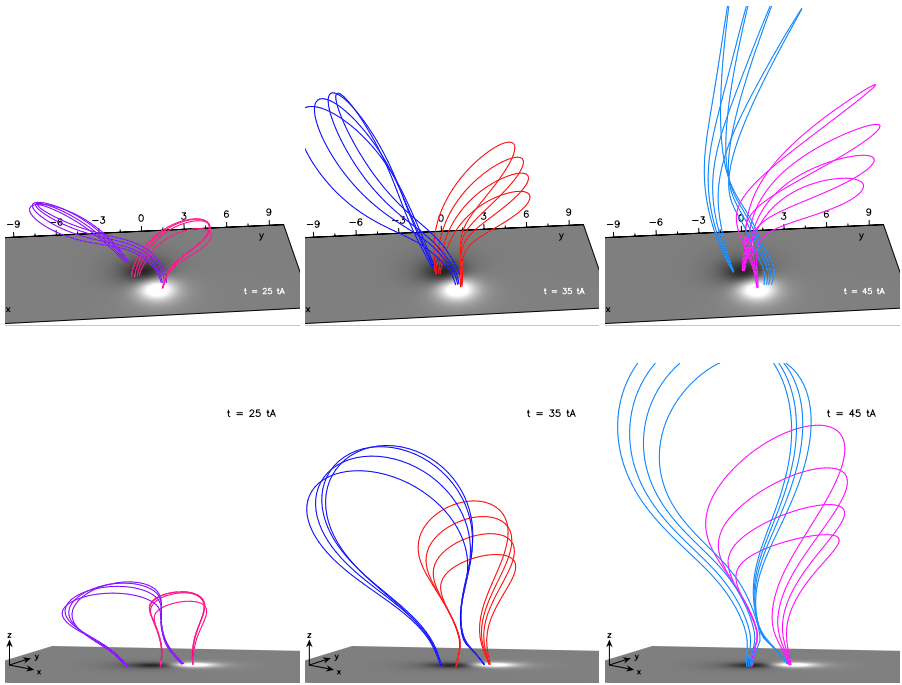}
     \caption{Three sets of reconnecting field lines at different times during the flux rope ejection. \textbf{Left:} side and front views at $t=25\ t_A$. \textbf{Middle:} a different set at $t=35\ t_A$. \textbf{Right:} Reconnecting field lines at $t = 45 t_A$. 
     }
     \label{fig13}
    \end{figure*}

\subsection{QSL evolution during the flux rope ejection}

The previous analyses concluded that $N$ does not significantly change on a time scale shorter than $t_A$.
However, the HFT as shown in the bottom row Fig. \ref{fig1} evolves throughout the flux rope ejection, which lasts a few tens of $t_A$, indicating that $N$ is likely to change during the entire solar eruptions.

Since the QSLs are displaced during the flux rope expansion, a reference parameter should be chosen for comparison at different times in the simulation. To quantify the evolution of the QSL in time, the maximum value of $N$ along the QSL crest could be measured as a reference. However, this study necessitates many iterations to follow the peak of $N$ in the whole 2D domain (at $z=0$) and does not necessarily relate to the previous calculations of $N$ for lines A and B. Moreover, it does not give the general evolution of the QSL at different locations. Therefore, we calculated the \Npeak\ values at different  locations by making several $x$-cuts of the $N$ profile, from $y=-2$ to $y=-6$. Then the general tendency of the \Npeak\ evolution within similar regions can be observed.

The logarithm of the peak value \Npeak\ is reported in Fig.~\ref{fig12}, where two domains, for $t \leq 22\ t_A$ and $t > 22\ t_A$, are defined. In the first domain with $t \leq 22\ t_A$, there is no clear time evolution of \Npeak. Note that the lack of data for $y=-5,-6$ at earlier times is due to the portion of the studied QSL that does not extend farther than $y=-4$.
In the second domain, however, ($t > 22\ t_A$), there is a clear increase in the \Npeak\ value as time advances, with an exponential trend. Interestingly, as time evolves, the \Npeak\ values tend to a similar value that itself tends to infinity.

The clear transition between the two domains can be understood from different physical processes at work. Eventually, the two HFT branches show an increase in the squashing degree value as well as a thinning of the QSLs (Fig. \ref{fig1}). To readily understand the changes in the configuration, different sets of reconnecting field lines are presented at three different times in the simulation in Fig. \ref{fig13}.
Front and side views are presented to clearly show the vertical expansion of the flux rope during its ejection.
This vertical stretching gradually leads to planar field lines, and reconnection then takes place similarly as in 2D models, as can be seen at $t=45\ t_A$ (right panels, see also \citetalias{Aulanier12}, Figure 5). QSLs in such a system evolve toward separatrices, so that a fast, exponential-like evolution of $N$ can be expected and is confirmed in Fig. \ref{fig12}. However, this explanation is only valid at later times in the simulation, \ie, for $t > 22\ t_A$.

The question remains why at $t \leq 22\ t_A$ there is no clear \Npeak\ behavior. Different physical explanations can be given: in the early stage of the simulation, even though the shearing motion at the photosphere is stopped, residual forces from the simulation at $t = 0$ still act on the magnetic field while the instability develops. These forces influence the dynamics of the QSLs related with the initial magnetic configuration. Therefore, the QSLs evolution cannot only be associated with the instability leading to the flux rope expansion and associated reconnection. Moreover, the differential magnetic shear around the flux rope created by the differential velocity shear at $t<0$, as discussed in \citetalias{Aulanier12}, participates in the QSLs found in the early stages of the numerical simulation.
Finally, note that the resistive diffusion coefficient $\eta$ is doubled at $t = 16\ t_A$ (to ensure numerical stability). Increasing the resistivity impacts the reconnection dynamics of magnetic field lines and can explain the fluctuations of \Npeak($t$) between $t=10\ t_A$ and $t=20\ t_A$. Eventually, the flux rope expansion and the reconnection of field lines become the main driver of QSLs dynamics, as is shown in the time evolution of \Npeak\ for $t > 22\ t_A$. Then, doubling the value of the resistivity again at $t = 30\ t_A$ does not affect the exponential-like evolution of \Npeak.

Since \vslip\ is correlated with $N$ during a short $\Delta t$ within which slip-running reconnection takes place, the present result suggests that \vpeak\ probably also evolves during the flux rope ejection. This has been qualitatively verified in the present simulation by investigating several \vslip\ profiles for different sets of field lines and at different times. However, since a very high time resolution is required to obtain an accurate \vpeak\ value, quantifying \vpeak$=f(t)$ would require a rather cumbersome analysis. Furthermore, since QSLs evolve toward separatrices, this means that \vslip$\rightarrow \infty$, so that quantifying \vpeak\ for $t > 22\ t_A$ is harder.


\section{Summary}
\label{secsum}

We aimed at understanding the main characteristics of 3D reconnection for the formation of flare loops during an eruptive flare and in the absence of a magnetic null point to extend the standard solar eruptive flare model from 2D to 3D. To do this, we analyzed a magnetic configuration modeling a simple bipolar active region, with quasi-separatrix layers (regions of high magnetic connectivity distortion) and without a null point. A 3D MHD numerical simulation was used to model the expansion of a flux rope during an eruptive flare event. QSLs lead to 3D reconnection dynamics that form both the flux rope envelope and the flare loops.

The core of the QSL is an HFT that is located in the corona, below the flux rope \citep{Titov07}. This HFT can locally be regarded as an X-point, with a guide field that is not invariant by translation, and that connects down to the photosphere. 
We showed that the location and the evolution of the QSLs are strongly correlated with high current build-up regions where 3D reconnection takes place.
Hence, the HFT is surrounded by a coronal current layer in which the flare reconnection takes place. The photospheric footprints of the QSL form a double $J$-shaped pattern \citep{Demoulin96b, Savcheva12}. The hooked parts of the $J$s surround the legs of the flux rope, and comprise both direct and return currents. The straight parts of the $J$s are the sites of direct currents only. They correspond to flare ribbons \citep{Qiu07}, and join the coronal HFT by a cusp-shaped feature.  

We demonstrated that the dynamics of the formation of flare loops and the flux rope via 3D reconnection is associated with the flipping motion of field lines. This flipping motion corresponds to successive reconnections that are a signature of QSL reconnection.
Investigating the slipping motion speed \vslip, we showed that the apparent motion of magnetic field lines is super-alfv\'enic, which is why we named this dynamics a slip-running reconnection phenomenon.

Investigating \vslip\ in great detail for different field lines at a given time, we also showed that its time profile is similar to the spatial profile of the field line mapping norm $N$, a parameter used to quantify QSLs in the reconnection time interval. Furthermore, we statistically showed a strict linear correlation between the two quantities. Therefore, $N$ was found to be a much better proxy than the squashing degree $Q$, or even \Qroot, to relate slipping reconnection motions with QSL properties. Since the vertical magnetic field magnitudes at the two footpoints of reconnecting flux tubes are different in typically asymmetric solar active regions, one can expect different field line motion velocities in each polarity.

Next, we showed that the QSL parameter $N$ evolves with an exponential-like increase in the simulation when the flux rope ejection becomes the main driver of QSL dynamics. This fast increase is driven by the pre-reconnection magnetic field lines eventually becoming more 2D. This is a consequence of the expansion of the flux rope, which stretches the whole structure, and of the reconnection of more potential field lines in the outer part of the magnetic configuration. This results in thinner QSLs with higher $N$ values, so evolving toward separatrices, as demonstrated in vertical cuts of the HFT at different times. 

To summarize, we have shown that the slipping motion of field lines in 3D reconnection can be characterized by the norm $N$ of the field line mapping. The slipping motion \vslip\ and the mapping norm $N$ are strictly and linearly correlated, with $v_{\mathrm{slip}} (t)\approx v_{\mathrm{QSL}} N(t)$, where $v_{\mathrm{QSL}}$ is the displacement velocity of the QSL at the photosphere.

\section{Conclusion}
\label{secccl}

The present numerical results, considered together with those of the first two papers of the present series \citep{Aulanier12,Aulanier13}, constitute 3D extensions to the magnetic field structure and dynamics in the standard CSHKP model for solar eruptive flares.
These extensions primarily consist of introducing the 3D physics of magnetic  reconnection at a QSL that surrounds an erupting flux rope that is initially weakly twisted, with a pair of $J$-shaped loops, and nearly aligned with
the PIL.

The global nature of the flare reconnection is slip-running: individual field lines slip at super-Alfv\'enic speeds within slowly moving plasma \citep{Aulanier06}. For a given reconnection rate, these very high apparent field line speeds are determined by the thickness of the QSL. At a given time during the eruption, the magnitude of the slipping speed of an elementary flux tube is proportional to the norm $N$ \citep{Priest95} of the QSL at its footpoint.
During the eruption, the reconnection taking place in the coronal current layer leads to the displacement of the QSL footprints away from the PIL, just like flare ribbons do. These QSL footprints then swipe the footpoints of the field lines while they reconnect at the HFT.
Then, the acceleration of the apparent motion of a flux tube footpoint is determined by the spatial profile of $N$ across the QSL footprint.

The slip-running flare reconnection forms flare loops that undergo a strong-to-weak shear transition, as well as a broad and more-and-more twisted envelope around the initially weakly twisted erupting flux rope (see \citetalias{Aulanier12}). These time-evolutions come from the combined effects of the reconnection-driven transfer of magnetic shear from the pre-erupting coronal arcades into the flare loops (see also \citealt{Schmieder96,Su06}) and around the erupting rope, and of the vertical stretching of the whole system. The latter leads to a diminishing of the spatially non-uniform guide field. This leads to the formation of nearly potential flare loops and a highly twisted envelope later in time. This process is associated with a sharp amplification in time of the norm $N$ of the QSL by several orders of magnitudes, which in turns leads to a strong increase in the slip-running speeds.
Eventually, the reconnection tends to become two-dimensional, which
incidentally leads to the recovering of the CSHKP flare model late in the eruption.

The previous successful associations of this numerical model to various observed features and to the estimation of flare energies suggest that the present extensions to the 2D standard flare model are generic. Moreover, even though the flare reconnection in the present numerical model is a result of the expansion of a torus-unstable flux rope \citep{Aulanier10}, it is arguable that these extensions will hold in other simulations in which the flare reconnection is driven by different mechanisms.
Indeed, these extensions rely on the existence of an erupting twisted flux rope. Such a magnetic structure is detected in-situ as magnetic clouds even for large impact parameters \citep{Demoulin13}. Moreover, including the magnetic could-like ICMEs, evidence of flux ropes in in-situ data accounts for at least two-thirds of observed ICMEs (see \citealt{Zurbuchen06} and \citealt{Riley12}).


\begin{acknowledgements}
The MHD calculations were performed on the quadri-core bi-Xeon computers of the Cluster of 
the Division Informatique de l'Observatoire de Paris (DIO). 
The work of M.J. is funded by a contract from the AXA Research Fund.  
\end{acknowledgements}


\bibliographystyle{aa} 
\bibliography{janvierREF}  
\IfFileExists{\jobname.bbl}{} {\typeout{}
\typeout{***************************************************************}
\typeout{***************************************************************}
\typeout{** Please run "bibtex \jobname" to obtain the bibliography} 
\typeout{** and re-run "latex \jobname" twice to fix references} 
\typeout{***************************************************************}
\typeout{***************************************************************}
\typeout{}}

\end{document}